\documentclass[aps,pre,twocolumn,groupedaddress,showpacs,superscriptaddress]{revtex4-1}

\usepackage{graphicx}
\usepackage{dcolumn}
\usepackage{bm}
\usepackage{units}
\usepackage{float,amsmath,verbatim,amssymb,stmaryrd}
\usepackage{epsfig}
\usepackage{color}

\usepackage[colorinlistoftodos,textsize=tiny]{todonotes}

\begin{document}

\preprint{APS/123-QED}

\title{Numerical heating in particle-in-cell simulations with Monte Carlo binary collisions}

\author{E. P. Alves}
\email{epalves@slac.stanford.edu}
\affiliation{
High Energy Density Science Division, SLAC National Accelerator Laboratory, Menlo Park, CA 94025, USA
}

\author{W. B. Mori}
\affiliation{Department of Physics and Astronomy, University of California Los Angeles, Los Angeles, CA 90095, USA}

\author{F. Fiuza}
\email{fiuza@slac.stanford.edu}
\affiliation{
High Energy Density Science Division, SLAC National Accelerator Laboratory, Menlo Park, CA 94025, USA
}

\begin{abstract}
The binary Monte Carlo (MC) collision algorithm is a standard and robust method to include binary Coulomb collision effects in particle-in-cell (PIC) simulations of plasmas. Here, we show that the coupling between PIC and MC algorithms can give rise to (nonphysical) numerical heating of the system, that significantly exceeds that observed when these algorithms operate independently. We argue that this deleterious effect results from an inconsistency between the particle motion associated with MC-collisions and the work performed by the collective electromagnetic field on the PIC grid. This inconsistency manifests as the (artificial) stochastic production of electromagnetic energy, which ultimately heats the plasma particles. The MC-induced numerical heating can significantly impact the evolution of the simulated system for long simulation times ($\gtrsim 10^3$ collision periods, for typical numerical parameters). We describe the source of the MC-induced numerical heating analytically and discuss strategies to minimize it.
\end{abstract}

\maketitle

\section{Introduction}
\label{S:1}

The particle-in-cell (PIC) method \cite{Dawson1983,Hockney1988,Birdsall1985} is a robust and versatile simulation technique to capture kinetic plasma effects in large scale systems and in multiple dimensions. It has been successfully used in a vast number of areas in plasma physics research, including plasma-based accelerators \cite{TajimaDawson1979, Litos2014}, intense laser-plasma interactions \cite{Fiuza:2012kd, Ridgers2012}, plasma instabilities \cite{Silva:2003tq,Alves:2012vb,Huntington:2015ko, Alves2019}, and astrophysical plasma phenomena \cite{Alves2018,Spitkovsky:2008wo,Guo:2014gs}. In this method, the coupling between the motion of a collection of plasma particles and their electromagnetic field is described self-consistently. The plasma particles have a finite-size associated with the deposition of the particle current/charge densities on the simulation grid, which is used to update the electromagnetic fields at the grid vertices, by solving the field equations in a discretized form. In the case of electromagnetic PIC codes, the field equations are Maxwell's equations. While the long-range (collective) electromagnetic field is well described, this method distorts and reduces (smooths) the short-range (inter-particle) electromagnetic fields thus allowing studying kinetic physics with few particles than in a real plasma \cite{Dawson1983,Hockney1988,Birdsall1985}. As a result, the standard PIC method does not quantitatively capture collisional plasma behavior associated with particle-particle interactions \cite{note1}. It does however, capture collisions between the finite size particles which has a modified collision operator from a real plasma.

In physical regimes where collisions between plasma particles are important, a Monte Carlo (MC) procedure is commonly adopted to include the statistical effects of collisions on the plasma dynamics in a quantitatively accurate manner. When doing this it is imperative that the effects from the collisions between the finite size particles be reduced to sufficiently low levels. 

There are two main MC methods that have been applied to the PIC framework. The first is a grid-based method, where the moments of the collision field particles are defined on the simulation grid, and the simulated macro-particles are subject to drag and diffusion in velocity space according to Langevin equations. The Langevin equations are integrated via MC sampling. The drag and diffusion coefficients are functions of the moments of the velocity distribution of the field particles (defined on the grid) and the macro-particle velocities, and are chosen to satisfy the classical theory of screened Coulomb collisions \cite{Lemons:2009da, Cohen:2013if}. The second method is the binary collision method, where particles in close vicinity (within the same grid cell) are randomly paired and elastically scattered. The scattering angle is sampled via MC from a probability distribution function that describes the statistics of classical screened Coulomb collisions \cite{TakizukaAbe1977, Nanbu1997, Nanbu1998, Sentoku:2008iu, Perez:2012fy}. These methods are found in many of the PIC codes used by the plasma community (e.g. \cite{Fonseca2002,Fonseca:2008ib, Sentoku:2008iu,vpic,calder,smilei,epoch}) and have been used successfully to describe the dynamics of collisional plasmas in a variety of scenarios.

While there has been significant work on understanding the convergence properties of both these methods \cite{Wang:2008co,Cohen:cr} and extending their validity for a wider range of physical regimes \cite{Sentoku:2008iu, Peano:2009iy, Perez:2012fy, Higginson2017}, little focus has been given to the numerical and nonlinear coupling between these MC-based collision models and the PIC framework. This is in large part because most of the validation tests for these collision models (relaxation tests of temperature anisotropies, relative drifting species, etc) are performed isolated from the standard PIC algorithm, i.e. when the self-consistent long-range electric and magnetic fields calculated by the standard PIC algorithm are turned off \cite{TakizukaAbe1977,Sentoku:2008iu,Perez:2012fy}. In addition, the few test cases that do require the PIC-MC coupling (e.g. validation of resistive fields \cite{Perez:2012fy}) are performed for short simulation times, i.e. order $10$s of the inverse collision frequency. The effects of the PIC-MC coupling over long simulation times of $\gtrsim 10^3$ of the inverse collision frequency remain, to our knowledge, unexplored. Such long simulations are necessary, for instance, to model the dynamics of collisional shocks in multi-component plasmas for inertial confinement fusion (ICF), where kinetic effects like species separation at the shock front are expected to develop \cite{Amendt2011,Bellei:2014kq,Bellei:2014ks,Rinderknecht:2015et}.



In this work, we explore the numerical coupling between the binary MC collision algorithm \cite{TakizukaAbe1977} and the electromagnetic PIC framework \cite{Birdsall1985} in PIC-MC simulations. In particular, we find that PIC-MC simulations suffer from (non-physical) numerical heating that is significantly greater than when PIC and MC algorithms operate independently. We show that this numerical heating results from an inconsistency between the motion of the particles induced by the MC collisions and the work performed by the collective fields evaluated on the PIC grid. This inconsistency manifests as the artificial production of electromagnetic radiation that stochastically heats the neighboring plasma particles. We calculate the artificial production of electromagnetic energy for the special case of a uniform density plasma in thermal equilibrium, and derive scaling laws for the artificial heating rate with numerical parameters. We then verify our analytic scaling laws with the results of PIC-MC numerical experiments. These scalings are important to guide the choice of numerical parameters for simulating a given system and for guaranteeing its physical validity. In particular, as the PIC method is being pushed to model very large systems, often with $\gtrsim 10^6$ time steps, leveraging the continuous increase in computational power and the inherent advantages of the method for massively parallel computing, the numerical heating discussed in this work can become a significant concern and needs to be controlled.

This paper is organized as follows. We begin by discussing the coupling between the PIC and MC algorithms in Section \ref{S:2}. We show that the motion of the particles due to MC collisions are inconsistent with the work performed by the fields on the PIC grid, resulting in the artificial production of electromagnetic radiation. We further calculate the resulting heating rate for the special case of a plasma in thermal equilibrium, and determine how it scales with the numerical parameters of the simulation. In Section \ref{S:3}, we perform numerical experiments using 1D, 2D and 3D PIC-MC simulations and verify the results of our analytical estimates of the heating rate. Based on our findings, we discuss strategies to minimize the heating rate in Section \ref{S:4}, and present our conclusions in Section \ref{S:5}.

\section{PIC-MC coupling}
\label{S:2}

In the PIC method, charged particles interact with each other via self-consistently generated electric and magnetic fields. The simulation domain is discretized into a grid (hereafter, the PIC grid). Particles deposit their current/charge densities onto the PIC grid, and Maxwell's equations are used to self-consistently advance the electric and magnetic fields onto the same PIC grid. These fields are then interpolated to each of the individual particle positions to evaluate the Lorentz force and advance the particles to their new position. This procedure corresponds to the main loop (one time-step) of the PIC algorithm.

The particle current/charge density deposition on the PIC grid effectively acts as a low-pass filter, smoothing out the high-frequency (wavenumber) components of the current/charge densities \cite{Dawson1983}. The self-consistent electric and magnetic fields that are evaluated on the PIC grid therefore correspond to smooth collective fields, which correctly describe the long-range (collisionless) plasma interactions but mitigate short-range (collisional) interactions. The binary MC collision method aims to fill this gap. We note that there are short-range interactions between the plasma particles which are less severe than for point particles \cite{Okuda1970}. Therefore, the PIC method does include modified collisions that will cause a non-Maxwellian distribution function to relax towards a Maxwellian. The finite size particle collisions do not lead to numerical heating (which is caused by aliasing) as energy is conserved during the process. These collisions can be reduced by increasing the number of particles per cell.

In a pure binary MC collisional simulation, particles interact solely through binary collisions. Similar to the PIC method, the simulation domain is discretized into a grid (the MC grid, or collision grid). Particles within the same collision cell are randomly paired, and their momenta are scattered by a random angle which obeys some prescribed probability distribution function (PDF). When the colliding particles have equal numerical weights, this method conserves kinetic energy and momentum on each collision, and thus global energy and momentum conservation is also achieved. Extension to collisions between particles of different numerical weights has been treated in \cite{Nanbu1998,Sentoku:2008iu}. In this work, we restrict ourselves to collisions between particles of equal numerical weight for simplicity. The scattering angle PDF determines the collision statistics of the system and ultimately the macroscopic transport properties of the system. For plasmas, the standard scattering angle PDF is such that Spitzer collision rates are recovered \cite{TakizukaAbe1977}. However, numerous works have proposed modifications to the scattering angle PDF to extend the validity of the macroscopic collision rates to broader physical regimes of plasma density and temperature \cite{TakizukaAbe1977,Nanbu1997,Nanbu1998,Sentoku:2008iu,Peano:2009iy,Perez:2012fy,Turrell2015,Higginson2017}.

\begin{figure*}[t!]
\begin{center}
\includegraphics[width=\textwidth]{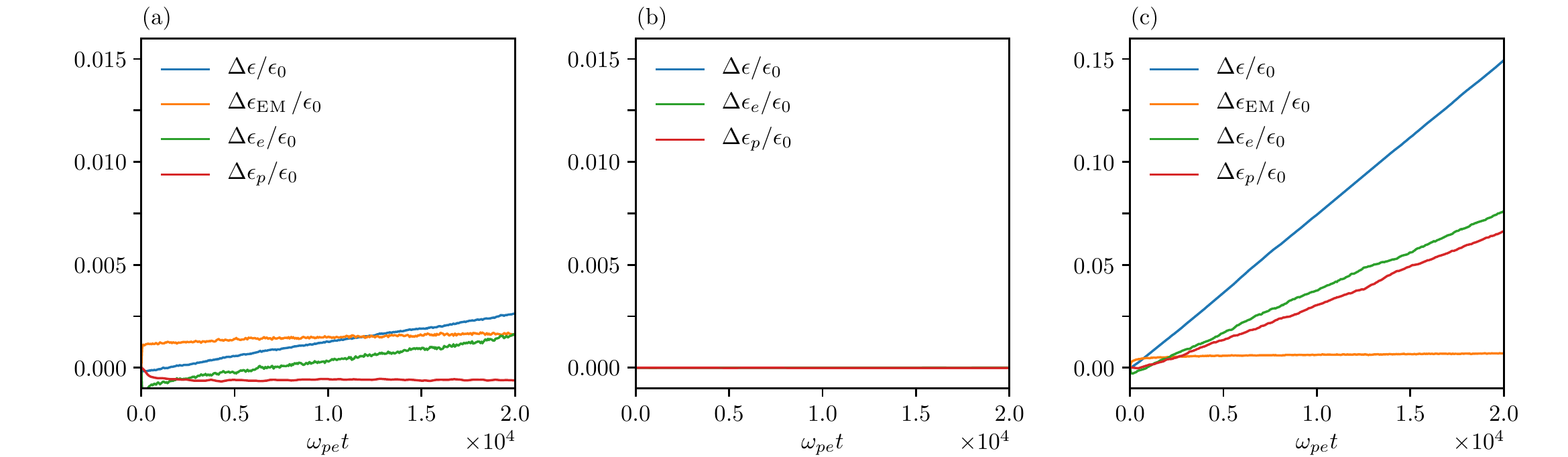}
\caption{Comparison between pure PIC (a), pure MC (b) and PIC-MC (c) simulations of a 1D uniform electron-proton plasma in thermal equilibrium. Total energy conservation ($\Delta \epsilon/\epsilon_0$), change in electron ($\Delta \epsilon_{e}/\epsilon_0$) and proton ($\Delta \epsilon_{p}/\epsilon_0$) kinetic energies are shown for each case. The change in electromagnetic energy ($\Delta \epsilon_{EM}/\epsilon_0$) is shown only for the PIC and PIC-MC simulations, since the self-consistent electromagnetic field is turned off in the pure MC case. Note that the range of the vertical scale of the PIC-MC case is $10$ times larger than the pure PIC and pure MC cases.}
\label{fig1}
\end{center}
\end{figure*}

The PIC and binary MC collision algorithms can thus be naturally combined to simultaneously capture both long-range (collective) and short-range (particle-particle) interactions (so long as enough particles are used to reduce the finite size collisions to low enough levels). The collision-grid is usually made to coincide with the PIC grid, and particles within the same cell are sequentially pushed according to i) the long-range electric and magnetic fields, and  ii) are randomly paired and scattered according to the local collision statistics. We find, however, that the coupling between PIC and MC algorithms can lead to unphysical effects that are enhanced compared to when these two algorithms operate independently.  While the kinematics of the MC collisional interactions between particle pairs preserves energy and momentum, the stochastic velocity changes ($\mathbf{\delta v}_{\alpha\beta}$) result in particle displacements that are inconsistent with the work performed by the collective fields registered on the PIC grid. This inconsistency manifests as a stochastic error in the field values on the grid, which ultimately leads to the artificial stochastic heating of the system and deterioration of global energy conservation.


This effect is illustrated in Figure~\ref{fig1}, where the energy conservation results for pure PIC (a), pure MC (b) and PIC-MC (c) simulations of a one-dimensional (1D) electron-proton plasma in thermal equilibrium are presented as a function of time. 
The initial plasma temperature is $T=100 \mathrm{eV}$ and the number density is $10^{24} \mathrm{cm}^{-3}$, which are typical conditions that arise in intense laser-solid interactions and ICF research \cite{Casanova1991,Dervieux2015}. For these conditions, the characteristic electron-ion collision frequency ($\nu_{ei}$, defined in Section \ref{S:3}) is $\nu_{ei} \simeq 0.3 \omega_{pe}$, where $\omega_{pe}$ is the electron plasma frequency. We used a grid resolution of $\Delta x = 0.25 c/\omega_{pe} \simeq 18 \lambda_D$ (where $c/\omega_{pe}$ and $\lambda_D$ are the plasma skin depth and Debye length, respectively), $100$ particles per cell per species, and a time step close to the Courant stability condition for an explicit fully electromagnetic field solver; note that by temporally resolving $\omega_{pe}$, $\nu_{ei}$ is automatically well resolved since $\nu_{ei}<\omega_{pe}$ in all cases considered in this work. Both physical and numerical parameters are kept fixed for the three simulation cases. Figure~\ref{fig1}-a reveals the innate finite numerical heating of PIC codes as seen by the artificial increase in the electron kinetic energy. This unwanted numerical effect is associated with the presence of aliasing modes in the electromagnetic field, but can be kept at a tolerable level by an appropriate choice of numerical parameters \cite{Hockney1988,Birdsall1985}. In our case, despite under resolving the Debye length (as is often the case in practical numerical simulations of intense laser-plasma interactions at high densities) we have used fourth order particle shapes to minimize the unphysical heating. This effect is explained by the presence of an artificial stochastic electric field (with zero mean and finite variance) that accounts for the stochastic errors inherent to the PIC numerics \cite{Hockney1988}. It has been shown that such a stochastic electric field leads to a linear increase in the average particle energy at a rate that is inversely proportional to the particle mass, which is consistent with the curves for the electron and proton energies in Figure~\ref{fig1}-a. For this reason, this innate numerical heating effect of PIC codes is also known as stochastic numerical heating. The results of the pure binary MC collisional simulation are presented in Figure~\ref{fig1}-b. Note that $\Delta \epsilon_{EM}/\epsilon_0$ is not shown for this case since the self-consistent electromagnetic fields are turned off. Both electron and proton species have equal numerical weights and hence the binary MC collisional interactions conserve kinetic energy and momentum both locally and globally. Unfortunately, when both PIC and MC algorithms are coupled a significant increase in the heating rate is found ($\simeq 60 \times$ the heating rate observed in PIC alone, for the chosen parameters), as observed in Figure~\ref{fig1}-c. In this case, we also observe the protons following the electron heating, due to the fast equilibration with the electrons compared with the heating time ($\nu_{ei} \gg \Gamma_\mathrm{MC}$, where $\Gamma_\mathrm{MC}$ is the MC-induced heating rate defined as $\Gamma_\mathrm{MC} \equiv (\Delta \epsilon/\epsilon_0)/\Delta t$). These numerical experiments suggest that the MC collision kinematics are not directly introducing errors into the particle kinetic energies, but may be introducing errors in the collective electromagnetic fields through modifications in the particle displacements and current densities.


\begin{figure*}[t!]
\begin{center}
\includegraphics[width=1.0\textwidth]{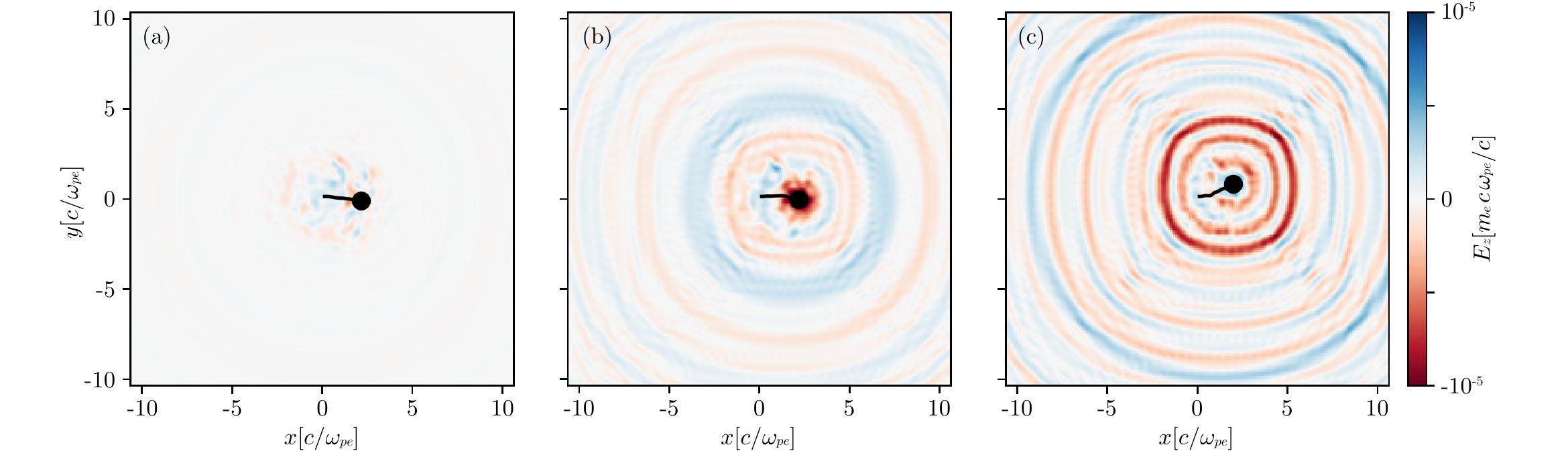}
\caption{Self-consistent radiation emission by a single electron undergoing MC collisions with plasma ions in 2D PIC-MC simulations. The color maps show the patterns of the out-of-plane component of the electric field ($E_z$), representing the electromagnetic emission of a single electron. The electron is traveling from left to right with velocity $v = \langle v \rangle = \sqrt{8/\pi} v_{the}$, where $v_{the}$ is the thermal electron velocity, and is represented by the black circle; its trajectory is traced by the black line. Panel (a) corresponds to the fields produced by the electron in a pure PIC simulation, and panels (b) and (c) correspond to the fields produced in PIC-MC simulations for $\nu_{ei} = 0.04\omega_{pe}$ and $\nu_{ei} = 0.2\omega_{pe}$, respectively.}
\label{fig2}
\end{center}
\end{figure*}

The effects of the PIC-MC coupling are further elucidated in the following numerical experiments. We use a subtraction technique \cite{Decyk:1987vt} to inspect the fields produced by a single electron in a PIC-MC simulation, revealing the effects of the PIC-MC coupling at the most basic level. This is achieved by performing a pair of 2D simulations of an electron-ion plasma in thermal equilibrium. The first is a pure PIC simulation, and the second is a PIC-MC simulation of precisely the same system (the initial particle positions and momenta are the same) with the addition of a single test electron that drifts from left to right in the plasma with velocity close to the thermal velocity. In the PIC-MC simulation, only the test-charge undergoes MC collisions with the background ions; the background electrons and ions do not interact via MC collisions. Hence, subtracting the electromagnetic field distributions between the two simulations yields the fields associated with a single thermal electron propagating in the background plasma while undergoing MC collisions with the background ions.

The out-of-plane component of the electric field ($E_z$) produced by the test electron is presented in Figure \ref{fig2}. This field component corresponds to a purely electromagnetic component of the radiation emitted by the test electron. Figures \ref{fig2} (b) and (c) reveal the radiation emitted as a result of the MC collisions for different collision frequencies ($\nu_{ei} = 0.04\omega_{pe}$ and $\nu_{ei} = 0.2\omega_{pe}$, respectively). The amplitude of the emitted radiation is observed to increase with the collision frequency, which is consistent with the increasing amplitude in the collisional velocity changes experienced by the electron; we have verified that the radiated power is consistent with Larmor's formula. This radiation is almost absent in Figure \ref{fig2} (a), which corresponds to a case where the collision frequency is zero, i.e. the case of a pure PIC simulation; the faint emission in this case results from numerical PIC collisions, occurring when the test electron scatters off of field fluctuations on the PIC grid. While it is expected that the PIC algorithm captures the electromagnetic radiation associated with the MC collisions through their effect on the current density deposited on the grid, this radiation violates the consistency of the PIC-MC algorithm. This is because the electromagnetic emission is the result of a force (associated with the MC collision) that is inconsistent with the electromagnetic field registered on the PIC grid. This error impacts the balanced energy exchange between particles and the electromagnetic fields on the grid that ensure global energy conservation. This radiation can therefore be seen as the artificial production of electromagnetic energy associated with MC collisions in PIC-MC simulations. It is this energy that is subsequently absorbed by the neighboring plasma particles, and ultimately increases the temperature of the system. 

Based on the above interpretation, we may estimate the numerical heating rate observed in PIC-MC simulations by calculating the rate of production of electromagnetic energy due to the MC-collisional interactions. For a closed (or periodic) and spatially uniform system, the averaged change in electromagnetic energy density after one time-step in a PIC-MC simulation is given by:

\begin{equation}
  \begin{split}
	\left\langle \frac{\epsilon_{EM,g}^{n+1}-\epsilon_{EM,g}^{n}}{\Delta t} \right\rangle
    &= -\left\langle
    \mathbf{J}_g^{n+1/2} \cdot \frac{\mathbf{E}_g^n + \mathbf{E}_g^{n+1}}{2}
    \right\rangle,
  \end{split}
\label{eq:deltawem}
\end{equation}

where we have assumed a standard leapfrog integration scheme commonly used in explicit PIC codes. The superscript $n$ denotes integer time step and the subscript $g$ denotes the grid cell index. The electromagnetic energy density at grid cell $g$ is given by $\epsilon_{EM,g}^{n} = (\mathbf{E}_g^n \cdot \mathbf{E}_g^n + \mathbf{B}_g^n \cdot \mathbf{B}_g^n)/8\pi$ (in CGS units), where $\mathbf{E}$ and  $\mathbf{B}$ are the electric and magnetic fields,  and $\mathbf{J}$ is the current density. The net Poynting flux is zero for a closed (or periodic) system.

We can explicitly decompose the total current density in the form $\mathbf{J} = \mathbf{J}_\mathrm{PIC} + \mathbf{\delta J}_\mathrm{MC}$, where $\mathbf{J}_\mathrm{PIC}$ is the current density associated with the pre-collision particle velocities, and $\mathbf{\delta J}_\mathrm{MC}$ is the current density associated with the MC collision velocity changes. For simplicity, we will refer to the latter as the collision current density, which can be expressed as

\begin{widetext}
\begin{equation}
  \begin{split}
   \mathbf{\delta J}_\mathrm{MC}\vert_g^{n+1/2}
    &= \sum_{(\alpha,\beta)} q_\alpha \mathbf{\delta v}_{\alpha,\beta}^{n+1/2} W_\alpha(\mathbf{x}_g-\mathbf{x}_\alpha)
    +q_\beta \mathbf{\delta v}_{\beta,\alpha}^{n+1/2} W_\beta(\mathbf{x}_g-\mathbf{x}_\beta) \\
    &= \sum_{(\alpha,\beta)} \left [
    \frac{q_\alpha}{m_\alpha} W_\alpha(\mathbf{x}_g-\mathbf{x}_\alpha)
    -\frac{q_\beta}{m_\beta} W_\beta(\mathbf{x}_g-\mathbf{x}_\beta)
    \right ]
     m_{\alpha\beta} \mathbf{\delta u}_{\alpha,\beta}^{n+1/2} \\
     &= \sum_{(\alpha,\beta)} \delta\rho_{\alpha\beta, g}~\mathbf{\delta u}_{\alpha,\beta}^{n+1/2},
  \end{split}
\label{eq:deltaj}
\end{equation}
\end{widetext}

where $q_i$ and $m_i$ are respectively the charge and mass of particle $i$; $W_i(\mathbf{x}_g-\mathbf{x}_i)$ is the numerical weight of particle $i$ at grid point $\mathbf{x}_g$; $\mathbf{\delta u}_{\alpha,\beta}$ is the change in the pre-collision relative velocity $\mathbf{u}_{\alpha,\beta} = \mathbf{v}_\alpha-\mathbf{v}_\beta$ between the pair of colliding particles $\alpha$ and $\beta$, and $m_{\alpha\beta} = m_\alpha m_\beta / (m_\alpha + m_\beta)$; $\mathbf{\delta v}_{\alpha,\beta}$ ($\mathbf{\delta v}_{\beta,\alpha}$) is the change in velocity of particle $\alpha$ ($\beta$) due to collision with $\beta$ ($\alpha$), and satisfies $m_\alpha \mathbf{\delta v}_{\alpha,\beta} = - m_\beta \mathbf{\delta v}_{\beta,\alpha} = m_{\alpha\beta}\mathbf{\delta u}_{\alpha,\beta}$. In the last line of Eq.~\ref{eq:deltaj}, we have written the collision current density as the sum of the contributions of effective particles with charge density $\delta\rho_{\alpha\beta} = m_{\alpha\beta} \left [ 
\frac{q_\alpha}{m_\alpha} W_\alpha(\mathbf{x}-\mathbf{x}_\alpha) -\frac{q_\beta}{m_\beta} W_\beta(\mathbf{x}-\mathbf{x}_\beta)
\right ]$, and effective velocity $\mathbf{\delta u}_{\alpha,\beta}$. It is this stochastic component of current density that is the source term for the production of radiation resulting from the MC collisions.

In a $1$D$1$V (one spatial dimension, one velocity component) system, the stochastic collision current density fluctuation at grid point $g$ will induce a change in the local electric field given by $\mathbf{\delta E}_g^{n+1/2} = \mathbf{E}_g^{n+1} - \mathbf{E}_g^{n} = - 4\pi \Delta t~\mathbf{\delta J}_\mathrm{MC}\vert_g^{n+1/2}$. Hence, the average rate of change in electromagnetic energy density due to stochastic current density fluctuations in a $1$D$1$V system is simply

\begin{widetext}
\begin{equation}
\left\langle \frac{\epsilon_{EM,g}^{n+1}-\epsilon_{EM,g}^{n}}{\Delta t} \right\rangle_\mathrm{MC}
    = -  \left\langle\mathbf{\delta J}_\mathrm{MC}\vert_g^{n+1/2} \cdot \mathbf{E}_g^{n} \right\rangle 
       +  2\pi\Delta t \left[ 
      \left\langle \mathbf{\delta J}_\mathrm{MC}\vert_g^{n+1/2} \right\rangle^2 + 
      \mathrm{Var}\left(\mathbf{\delta J}_\mathrm{MC}\vert_g^{n+1/2}\right)
     \right],     
\label{eq:deltawem2}
\end{equation}
\end{widetext}

where $\mathrm{Var}(X) = \langle X^2\rangle- \langle X\rangle^2$ denotes the variance of the random variable $X$. The first two terms are found to be zero if we consider current density fluctuations with zero average, and if the stochastic current density is uncorrelated with the local electric field of the previous time-step. Both these considerations are satisfied when we are dealing with a simple uniform plasma in thermal equilibrium. From Eq.~\ref{eq:deltawem2} it is clear that the presence of binary MC collisions gives rise to an artificial source term that is proportional to the variance of the stochastic collision current. The finite variance of the collision current derives from the finite particle statistics in a simulation cell, and is responsible for artificially injecting energy into the system and induce numerical plasma heating. We have verified through numerical simulations that in 2D and 3D the rate of increase of electromagnetic energy density continues to scale as $\Delta t~\mathrm{Var}\left(\mathbf{\delta J}_\mathrm{MC}\right)$, varying only in the proportionality constant by a factor of a few.

In the following, we will consider in more detail the numerical heating of a closed, uniform system in thermal equilibrium. Under these conditions,  we have that $\left\langle \mathbf{\delta J}_\mathrm{MC}\vert_g^{n+1/2}\right\rangle = \left\langle \mathbf{\delta J}_\mathrm{MC}\vert_g^{n+1/2} \cdot \mathbf{E}_g^{n} \right\rangle = 0$, allowing us to write the average rate of injection of electromagnetic energy density due to the stochastic collision current as:

\begin{eqnarray}
	\left\langle \frac{\epsilon_{EM,g}^{n+1}-\epsilon_{EM,g}^{n}}{\Delta t} \right\rangle_\mathrm{MC}
    \propto \Delta t~\mathrm{Var}\left(\mathbf{\delta J}_\mathrm{MC}\vert_g^{n+1/2}\right)
\label{eq:deltawem_thermal}
\end{eqnarray}

The electromagnetic energy produced by the MC collisions can be reabsorbed by the plasma particles via inverse bremsstrahlung (IB). As we will show in the next Section, the IB rate is significantly faster than the MC-induced heating rate for typical numerical parameters. We can therefore write that the change in plasma kinetic energy is given by




\begin{eqnarray}
	\left\langle \frac{\epsilon_{K,g}^{n+1}-\epsilon_{K,g}^{n}}{\Delta t} \right\rangle_\mathrm{MC}
    \propto \Delta t~\mathrm{Var}\left(\mathbf{\delta J}_\mathrm{MC}\vert_g^{n+1/2}\right),
\label{eq:deltaek_thermal}
\end{eqnarray}

where $\epsilon_K$ is the total kinetic energy density of the plasma at grid point $g$. Hence, for a uniform equilibrium plasma with initial thermal energy density ($\epsilon_0 = 2n_0~3/2k_B T_0$, where $k_B$ is the Boltzmann constant and $T_0$ is the initial plasma temperature), we define the MC-induced heating rate ($\Gamma_\mathrm{MC}$) as the average rate of increase of thermal energy density to the initial thermal energy density of the plasma:

\begin{eqnarray}
 \begin{split}
	\Gamma_\mathrm{MC} &\equiv 
	\frac{1}{\epsilon_{0}} \left\langle \frac{\epsilon_{K,g}^{n+1}-\epsilon_{K,g}^{n}}{\Delta t} \right\rangle_\mathrm{MC} \\
    &\propto
    \frac{ \Delta t~\mathrm{Var}\left(\mathbf{\delta J}_\mathrm{MC}\vert_g^{n+1/2}\right)}{n_0 k_B T_0}
\end{split}
\label{eq:gamma_mc}
\end{eqnarray}

From Eq.~\ref{eq:deltaj}, we find that $\mathrm{Var}(\mathbf{\delta J}_\mathrm{MC}\vert_g) = N_\mathrm{pairs} \langle \delta\rho_{\alpha\beta,g}^2 \rangle \langle \delta u_{\alpha\beta}^2\rangle$, where we have dropped the temporal superscripts to simplify the notation. The number $N_\mathrm{pairs}$ refers to the number of colliding particle pairs whose current will be deposited at a given grid point $g$. This number is proportional to the number of simulation particles per cell ($N_{ppc}$), and also depends on the order of the particle weighting scheme and the spatial dimensionality of the system. Hence, computing the variance of the stochastic collision current is reduced to computing the expected values for the random variables $\delta\rho_{\alpha\beta,g}^2$ and $\delta u_{\alpha\beta}^2$.


For simplicity, we consider collisions between species such that $m_\beta \gg m_\alpha$, as is the case for electron-ion collisions. In this limit, the random variable  $\delta\rho_{\alpha\beta, g}$ is reduced to $\delta\rho_{\alpha\beta, g} \simeq q_\alpha W_\alpha(\mathbf{x}_g -\mathbf{x}_\alpha)$. Given that the numerical weight of the particles will be proportional to $N_{ppc}$, we find that $\langle \delta\rho_{\alpha\beta,g}^2 \rangle \propto (q_\alpha n_0/N_{ppc})^2$.

The amplitude of the change in relative velocity between a pair of particles due to a binary collision can be written as  $\delta u_{\alpha\beta} = u_{\alpha\beta} \sqrt{2(1-\mathrm{cos}(\theta_{\alpha\beta}))}$, where $\theta_{\alpha\beta}$ is the scattering angle by which the relative velocity vector between particles $\alpha$ and $\beta$ is rotated after the collision. 
As mentioned before, the statistical distribution of the random scattering angle $\theta_{\alpha\beta}$  depends on the collision model used. A number of works have extended the original work of Takizuka \& Abe \cite{TakizukaAbe1977} to accommodate cumulative small-angle collisions \cite{Nanbu1997}, relativistic effects \cite{Peano:2009iy,Sentoku:2008iu,Perez:2012fy} and corrections at low-temperature and high-density plasma regimes \cite{Perez:2012fy}.
These extensions and corrections were achieved by modifying the statistical distribution of the random variable $\theta_{\alpha\beta}$, and therefore the stochastic current density and the associated heating rate will depend on the physical regime and the model used. Here, we will base our calculations on the original work by Takizuka \& Abe for non-relativistic Spitzer collisions.

From \cite{TakizukaAbe1977}, the probability density distribution of $\theta_{\alpha\beta}$ can be written in terms of another random variable $\delta$, such that $\delta u_{\alpha\beta} = 2~u_{\alpha\beta} \sqrt{\delta^2 / (1+\delta^2)}$, where $\delta$ is normally distributed with zero mean ($\langle \delta \rangle = 0$) and variance $\langle \delta^2 \rangle = 2\pi q_\alpha^2 q_\beta^2 n_L \lambda m_{\alpha\beta}^{-2}u_{\alpha\beta}^{-3} \Delta t$, where $\lambda$ is the Coulomb logarithm and $n_L = \mathrm{min}(n_\alpha,n_\beta)$. Moreover, the distribution of $u_{\alpha\beta}$ for two species in thermal equilibrium at temperature $T$ is found to be $f(u_{\alpha\beta}) = (m_{\alpha\beta}/2\pi T)^{3/2} 4\pi u_{\alpha\beta}^2$ $\mathrm{Exp}(-m_{\alpha\beta} u_{\alpha\beta}^2/2T)$. Based on these PDFs, and considering $\langle \delta^2 \rangle \ll 1$, we find that $\langle \delta u_{\alpha\beta}^2 \rangle \simeq (24/\pi)(T/m_{\alpha\beta}) \nu_{\alpha\beta}(T) \Delta t$, where $\nu_{\alpha\beta}(T) = 2\pi q_\alpha^2 q_\beta^2 n \lambda m_{\alpha\beta}^{-2}(T/m_{\alpha\beta})^{-3/2}$ is the characteristic collision frequency between species $\alpha$ and $\beta$ at temperature $T$.


We can now combine the calculated expectation values above to substitute in Eq.~\ref{eq:gamma_mc} and obtain the MC-induced heating rate for a closed electron-ion plasma system in thermal equilibrium:

\begin{eqnarray}
  \Gamma_\mathrm{MC}(T) = C_D \frac{T}{T_0} \frac{ \nu_{\alpha\beta}(T) (\omega_{pe}\Delta t)^2}{N_{ppc}},
\label{eq:gamma_mc2}
\end{eqnarray}

where $C_D$ is a constant that contains the effects of spatial dimensionality, particle shape, and spatial filters, which will be directly measured from PIC-MC simulations in the next Section. As expected, we find that the MC-induced heating rate is proportional to $\nu_{\alpha\beta}$, vanishing in the limit when the MC collisions are turned off ($\nu_{\alpha\beta}=0$). Eq.~\ref{eq:gamma_mc2} further reveals the intrinsic numerical (unphysical) character of this deleterious effect, through its dependence on the numerical parameters $N_{ppc}$ and $\Delta t$. The MC-induced heating rate diminishes with increasing number of particles per cell, as the collision-induced current fluctuations decrease with $1/N_{ppc}$. Moreover, the heating rate is also significantly reduced with decreasing time step as $\Delta t^2$, since smaller collisional velocity changes are obtained with smaller $\Delta t$, which in turn result in smaller collision-induced current fluctuations.

Note that Eq.~\ref{eq:gamma_mc2} expresses the MC-induced heating rate at the instantaneous plasma temperature $T$. This is because the fluctuations of the collision current density will change as the plasma is being artificially heated. It is convenient to explicitly separate out the instantaneous temperature dependence of $\Gamma_\mathrm{MC}(T)$ by writing

\begin{equation}
  \Gamma_\mathrm{MC}(T) = \sqrt{{\frac{T_0}{T}}}~\Gamma_\mathrm{MC}(T_0),
\label{eq:gamma_mcthree}
\end{equation}

where $\Gamma_\mathrm{MC}(T_0)$ is the MC-induced heating rate at the initial plasma temperature $T_0$. From Eq.~\ref{eq:gamma_mcthree}, we see that the $\Gamma_\mathrm{MC}(T)$ slowly decreases as the plasma is heated as $\sqrt{T_0/T}$. For instance, by the time the temperature of plasma has doubled, the instantaneous MC-induced heating rate decays only by a factor of $1/\sqrt{2}$. For most cases of interest, given that we want numerical heating to be minimized, we can therefore assume that the MC-induced heating rate remains approximately constant, and we will therefore use $\Gamma_\mathrm{MC}(T_0)$ as our main measure for the MC-induced heating rate in the PIC-MC simulations discussed in the next Section.

\section{Numerical experiments}
\label{S:3}
We have performed a series of numerical experiments to assess the MC-induced numerical heating in PIC-MC simulations using OSIRIS \cite{Fonseca2002,Fonseca:2008ib}. OSIRIS is a fully relativistic, fully electromagnetic PIC code, and is also equipped with a binary MC collisions module. For simplicity, in the following, we restrict our numerical tests to regimes where subrelativistic Spitzer collision rates are valid. In these regimes, all MC models are in agreement, and we will therefore limit our numerical experiments to the T\&A MC model for simplicity.

We begin by analyzing the MC-induced numerical heating of thermal equilibrium electron-ion plasma. For simplicity, we consider only collisions between electrons and ions, and neglect intra-species collisions. We simulate a periodic domain in 1D, 2D, and 3D, and explore MC-induced heating for different numerical parameters and collision frequencies. Ideally, one would choose a spatial resolution ($\Delta x$) that would match the Debye length ($\lambda_D$) of the plasma, but in most cases of dense plasma simulations this is impracticable, and the grid size is chosen to resolve the electron inertial length ($c/\omega_{pe}$) with a few points and high-order interpolation is used to guarantee good energy conservation \cite{Sentoku:2008iu}. In the following tests we choose $\Delta x = 0.25~c/\omega_{pe}$. As mentioned before, there is numerical heating intrinsic to the pure PIC algorithm itself \cite{Hockney1988,Birdsall1985}, which will also be present in the PIC-MC simulations. We denote the heating rate intrinsic to a pure PIC simulation by $\Gamma_\mathrm{PIC}$. We attempt to isolate the heating induced by the PIC-MC coupling from that intrinsic to a pure PIC simulation by choosing numerical parameters that keep $\Gamma_\mathrm{PIC}$ lower (less than half) than $\Gamma_\mathrm{PIC-MC}$ measured in PIC-MC simulations. We achieve this by using fourth-order particle shapes, and smoothing the current density field with a binomial compensated filter. For each set of numerical ($N_{ppc}$, $\Delta t$, $\Delta x$) and physical parameters (temperature $T_0$, density $n$, Coulomb logarithm $\lambda$), we perform both pure PIC and PIC-MC simulations, and subtract $\Gamma_\mathrm{PIC-MC} - \Gamma_\mathrm{PIC}$ to isolate $\Gamma_\mathrm{MC}(T_0)$ that stems from the PIC-MC coupling. Assuming that the sources of noise that give rise to $\Gamma_\mathrm{PIC}$ and $\Gamma_\mathrm{MC}(T_0)$ are uncorrelated and independent, then the difference $\Gamma_\mathrm{PIC-MC} - \Gamma_\mathrm{PIC}$ can provide an adequate measurement for $\Gamma_\mathrm{MC}(T_0)$.

\begin{figure}[t]
\begin{center}
\includegraphics[width=1.\linewidth]{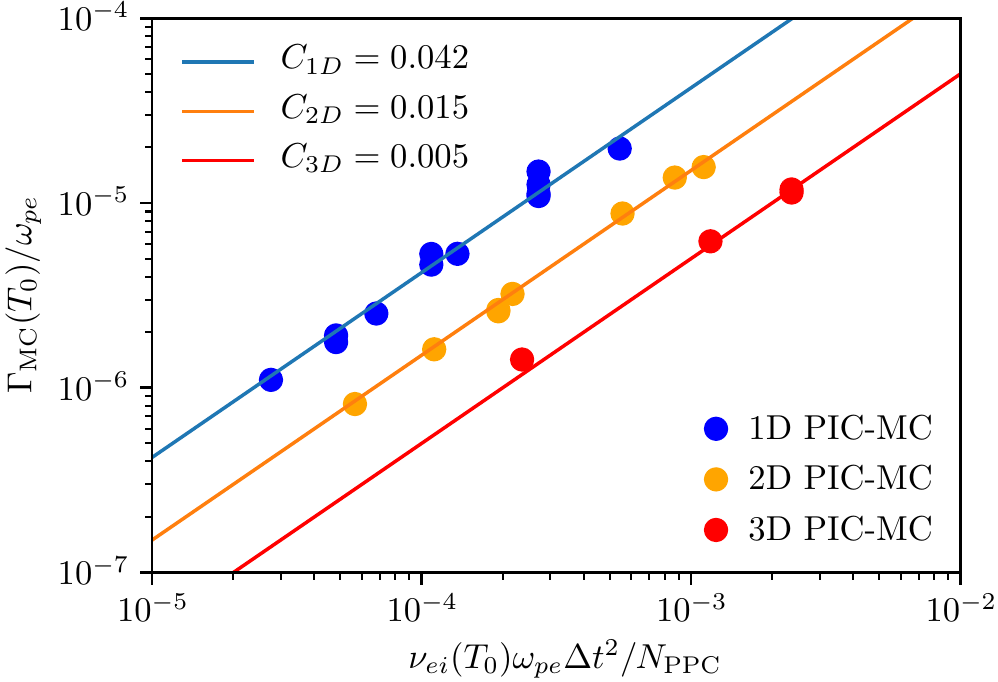}
\caption{MC-induced numerical heating in PIC-MC simulations of thermal equilibrium plasma. The points represent measurements from OSIRIS PIC-MC simulations, and the blue, orange and red colors represent 1D, 2D and 3D simulations, respectively. The measured heating rates are in agreement with the theoretical prediction of Eq.~\ref{eq:gamma_mc2} (solid lines). The heating rate coefficientes $C_D$ have been determined for 1D, 2D and 3D simulations. Note that the intrinsic numerical heating rate associated with the PIC algorithm has been subtracted from that of PIC-MC simulations to obtain these measurements.}
\label{fig3}
\end{center}
\end{figure}

The results are summarized in Figure~\ref{fig3}, and the measured heating rates are found to increase proportionally to $\nu_{ei}(T_0)(\omega_{pe}\Delta t)^2/N_{ppc}$ in agreement with Eq.~\ref{eq:gamma_mc2}. The heating rate constant $C_D$ is found to depend on the dimensionality of the simulation domain, and decrease by a factor of approximately 3 with each increasing dimension.
This decrease is related to the increasing number of cells that share the same grid point in higher dimensions. There are thus an increasing number of collision pairs that contribute to the collision current density at any given grid point, effectively reducing the fluctuations. 
Note that the collision frequency $\nu_{ei}$ was always well resolved in these simulations. The range of density and temperature parameters were varied such that $\nu_{ei}/\omega_{pe}<1/3$, and given the CFL stability condition for $\Delta t$ for an electromagnetic field solver, we have $\Delta t < \Delta x / c = 0.25 \omega_{pe}^{-1}$. Hence, $\nu_{ei} \Delta t < 1/12$ was held for all simulations.

The increase in the total energy in the simulations is observed primarily as particle heating, as seen in Fig. \ref{fig1}c). This is because the electromagnetic energy produced by the MC collisions is promptly reabsorbed by the plasma via IB. This is verified by comparing the IB and MC-induced heating rates. The collisional damping of electromagnetic waves of frequency $\omega$ by IB occurs at a rate $\Gamma_\mathrm{IB} = (\omega_{pe}/\omega)^2 (\nu_{ei}/2)$. Given that the highest frequency produced in the simulation is $\omega_\mathrm{max} = \pi/\Delta t$, the ratio between the IB and the numerical heating rates is
\begin{eqnarray}
  \frac{\Gamma_\mathrm{IB}}{\Gamma_\mathrm{MC}} = \frac{1}{C_D} \frac{N_{ppc}}{ (\omega \Delta t)^2} \geq \frac{N_{ppc}}{2 \pi^2 C_D}.
\label{eq:gamma_ib_mc}
\end{eqnarray}
We confirm that for the heating rate coefficients $C_D$ observed in Fig. \ref{fig3}, IB always dominates and thus the electromagnetic energy is expected to be rapidly absorbed by the plasma particles.

Our numerical heating scaling results provide a useful guide to choose the numerical parameters for a given PIC-MC simulation in order to ensure that the MC-induced numerical heating is kept at a tolerable level. Determining what level of numerical heating is tolerable will certainly be problem dependent, but if we are dealing with a closed physical system we can write this condition as $\tau_\mathrm{MC} \equiv 1/\Gamma_\mathrm{MC}(T_0) \gg T_\mathrm{sim}$, where $T_\mathrm{sim}$ is the simulated time of the system. This condition states that $T_\mathrm{sim}$ should be much smaller than the $\tau_\mathrm{MC}$, which corresponds to the time taken to approximately double the total energy of the system ($\Delta \epsilon/\epsilon_0 \simeq 1$). [If one takes into account the self-consistent temperature dependence of $\Gamma_\mathrm{MC}(T)$ as the system is being heated, then one finds that the energy of the system doubles after $\simeq1.22/\Gamma_\mathrm{MC}(T_0)$.] For the fourth order particle shapes and current density filter used in our simulations, and assuming a typical time step value of $\Delta t \simeq 0.2 \omega_{pe}^{-1}$, and $N_{ppc} \simeq 100$, $25$ and $8$ particles per cell in $1D$, $2D$,  and $3D$ simulations, we find that $\tau_{MC}\nu_{\alpha\beta} \simeq 5\times 10^4$. This shows that using typical numerical parameters $\Delta t$ and $N_{ppc}$, the binary MC collisional interactions artificially double the energy of the system after $\simeq 5 \times 10^4$ collision periods ($1/\nu_{\alpha\beta}$). 

\begin{figure}[t]
\begin{center}
\includegraphics[width=1\linewidth]{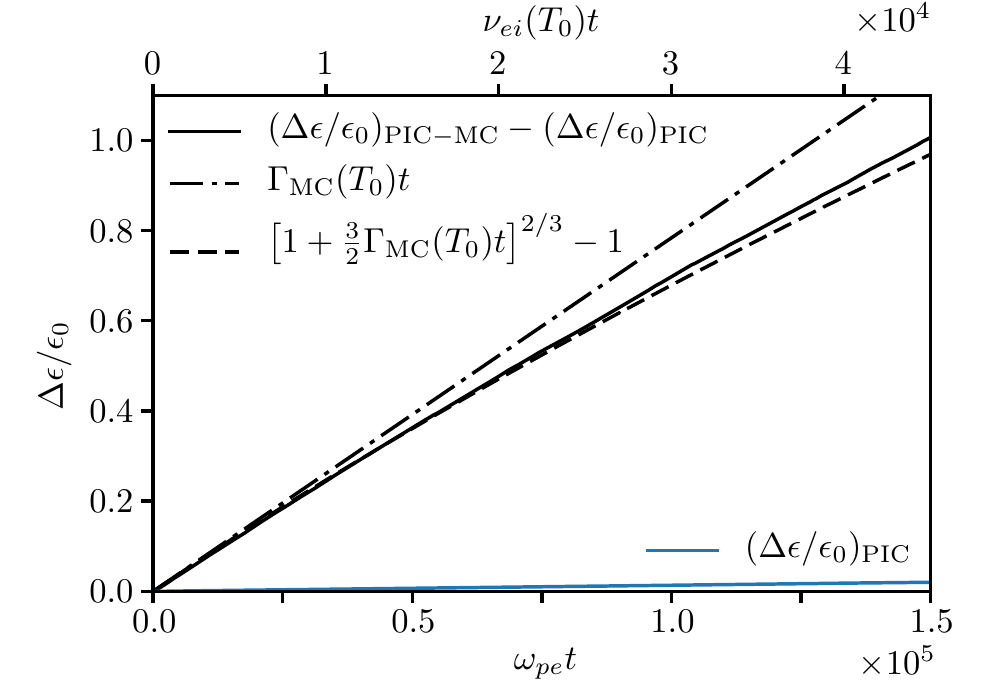}
\caption{Deterioration of energy conservation in a PIC-MC simulation due to MC-induced numerical heating (black solid curve), and comparison with theoretical predictions (dashed, and dash-dotted curves). The deterioration of energy conservation of a pure PIC simulation with the same numerical parameters is represented by the blue solid curve, and remains below $2\%$ for the simulated time. Note that the black solid curve corresponds to the difference in global energy variation between the PIC-MC and pure PIC simulations, in order to isolate the heating contribution associated with the PIC-MC coupling from the innate heating in PIC.}
\label{fig4}
\end{center}
\end{figure}

\begin{figure}[t]
\begin{center}
\includegraphics[width=1\linewidth]{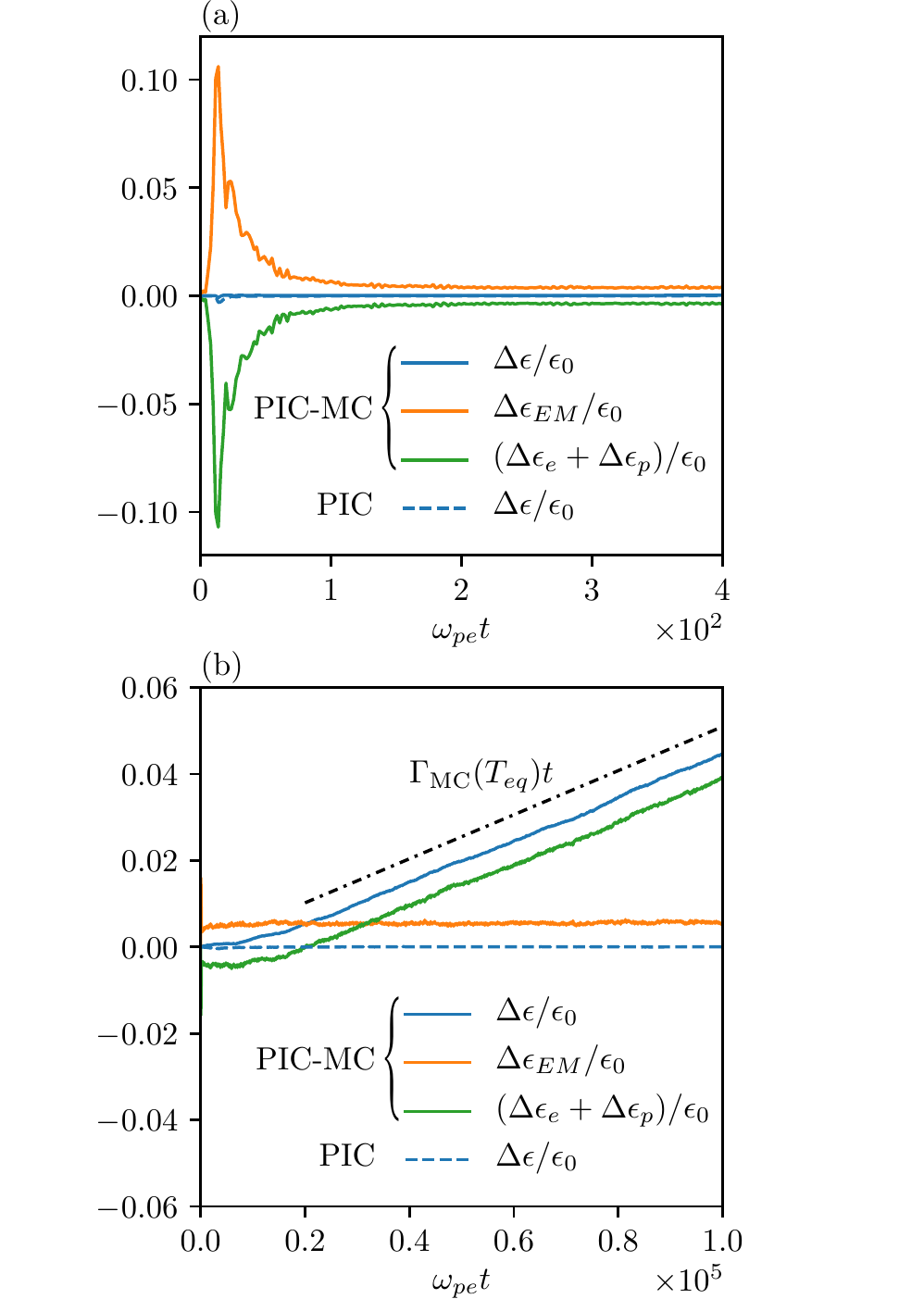}
\caption{MC-induced numerical heating in a two-stream unstable plasma. The early time evolution ($\omega_{pe}t<400$) of the system is shown in panel a), where the exchange between particle kinetic energy (solid green curve) and electrostatic energy (solid orange curve) due to the development of the instability is observed. During this time, identical global energy conservation curves are obtained for both PIC-MC (solid blue) and pure PIC (dashed blue) curves. This is because the effective MC-induced heating time is much larger than this time-scale. When the system relaxes to thermal equilibrium at around $\omega_{pe}t \simeq 2\times10^4$, the effective electron-ion collision frequency increases, and the MC-induced heating rate also increases. After this time we observe in panel b) the linear increase in particle energy, and associated deterioration of energy conservation at the rate given by $\Gamma_\mathrm{MC}(T_{eq})$ (dash-dotted black curve), where $T_{eq}$ is the temperature reached at thermal equilibrium.}
\label{fig5}
\end{center}
\end{figure}

\begin{figure*}[t!]
\begin{center}
\includegraphics[width=1.\linewidth]{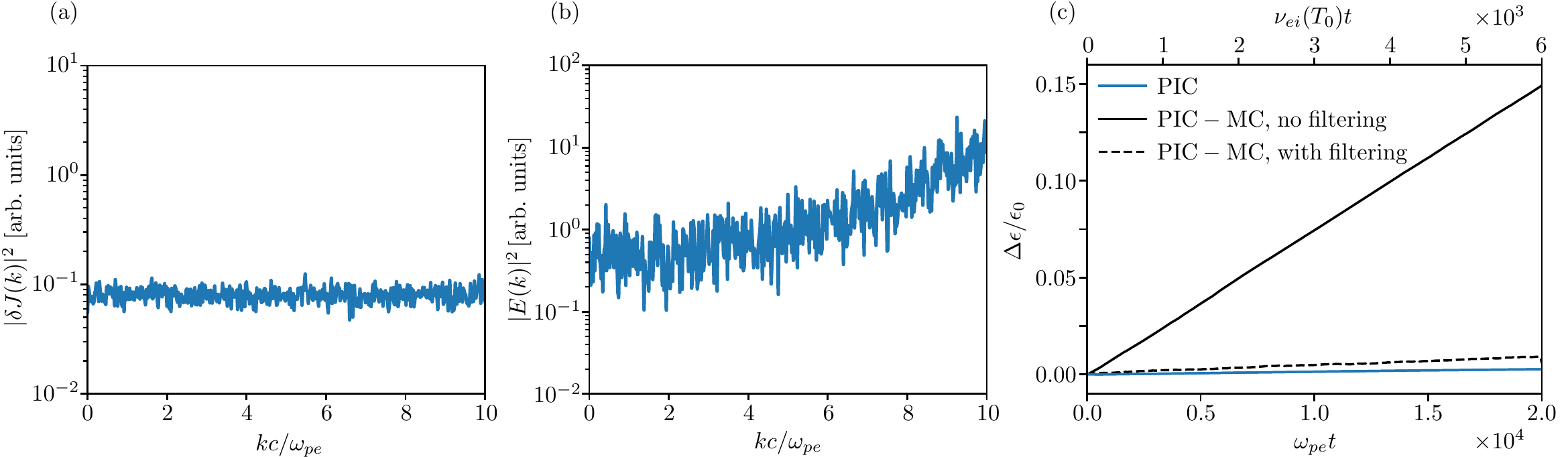}
\caption{
(a) Fourier spectrum of white noise current density fluctuations that mimic the effect of MC-collisions, and (b) Fourier spectrum of electric field fluctuations produced by the same noisy current density distribution. These spectra are obtained by integrating Maxwell's equations in 1D using a Finite Difference Time Domain solver in the presence of a noisy current density distribution. (c) Using a low pass filter on the electromagnetic fields at every $30\nu_{e}i^{-1}$ (every $500$ time steps), a large fraction of MC-induced electromagnetic fluctuations is suppressed, leading to a strong reduction in the MC-induced heating rate. This is illustrated for a 1D spatially uniform electron-proton plasma in thermal equilibrium with the same physical and numerical parameters used in Figure~\ref{fig1}. The solid (dashed) black and solid blue curves correspond to the PIC-MC simulation without (with) filtering on the fields, and a pure PIC simulation without MC collisions, respectively.
}
\label{fig6}
\end{center}
\end{figure*}

As an example, we present in Figure~\ref{fig4} the results of a 1D closed electron-ion plasma system with number density $n_0 = 10^{24} \mathrm{cm}^{-3}$ and in thermal equilibrium at $T_0=100\mathrm{eV}$, representing typical conditions in laser-solid interactions and inertial fusion relevant experiments. These conditions of temperature and density yield $\nu_{ei}\simeq 0.3\omega_{pe}$ (for $\lambda = 3$). The solid black curve in Figure~\ref{fig4} represents the deterioration of energy conservation in the PIC-MC simulation due to the MC-induced numerical heating, which is much greater than that associated with the innate numerical heating of the pure PIC simulation with the same physical and numerical parameters (blue solid curve). Note that we have subtracted the contribution of the pure PIC heating from the total variation in energy in the PIC-MC simulation, under the assumption that the PIC contribution remains the same in the PIC-MC case, and thus allowing us to isolate the MC-induced numerical heating. We see that at early times, the MC-induced heating is well characterized by the linear growth rate given by Eq.~\ref{eq:gamma_mc2} at the initial system temperature $T_0$ (dash-dotted line). The MC-induced heating rate is seen to slow down at later times when the system has significantly heated, as predicted by Eq.~\ref{eq:gamma_mcthree}. By integrating the MC-induced heating rate using the instantaneous temperature $T$ of the system, one obtains the expression represented by the dashed curve in Figure~\ref{fig4}, which is in good agreement with the behavior observed at late times in the PIC-MC simulation. We also verify that the energy of the system artificially doubles ($\Delta\epsilon/\epsilon_0=1$) after $\tau_\mathrm{MC}\nu_{ei}(T_0) \simeq 4.5\times10^4$. This corresponds to $\tau_\mathrm{MC}\omega_{pe} \simeq 1.5\times10^5$ or $\tau_\mathrm{MC} \simeq 6.5\times10^5 \Delta t$ for these parameters. For most cases of interest, numerical heating needs to be controlled to values $\lesssim 1\%$. In such cases, the MC-induced heating could limit the simulation time to only 500 collision periods ($1/\nu_{\alpha\beta}$).


Our analysis has been restricted to thermal equilibrium plasmas, and we have not yet addressed the MC-induced numerical heating in non-equilibrium configurations. In non-equilibrium plasmas, the finite variance of the MC collision current density will be modified by the local distribution function of the particles, and hence impact the numerical MC-induced heating rate. However, unless the system is being continuously driven out of equilibrium, the system will eventually relax to thermal equilibrium through the MC-collisions, and undergo the numerical heating at the rate predicted by Eq.~\ref{eq:gamma_mc2}. Since the equilibration occurs on a much shorter time-scale than the heating time-scale ($\Gamma_\mathrm{MC} \ll \nu_{\alpha\beta}$), as was verified by the results above, it is expected that the MC-induced heating will have a negligible effect during the transient non-equilibrium stage.

An example of such a scenario is presented in Figure~\ref{fig5}, where the results of a 1D simulation of a two-stream instability are shown. In this simulation, two populations of cold electrons are symmetrically counter-streaming in a background of stationary ions, and electron-ion collisions are turned on. Figure~\ref{fig5} a) shows the evolution of the system at early times ($\omega_{pe}t<400$), capturing the exponential development and decay of the two-stream instability as seen by the rise and decay of the electrostatic component of the electromagnetic energy density (orange curve). The global energy variation is represented by the solid blue curve, and is shown to remain at a very low level during this stage ($\Delta \epsilon/\epsilon_0 \simeq 10^{-4}$), and is identical to the level observed in the pure PIC simulation where collisions are turned off (dashed blue curve). At much later times around $\omega_{pe}t \simeq 2\times10^4$, the system has thermalized at temperature $T_{eq}$ (such that $2n_0~3/2k_B T_{eq} = \epsilon_0$). At around this time the electron-ion collision frequency increases, since the average relative speed between particles decreases, and we observe the steady numerical MC-induced heating rate given by $\Gamma_\mathrm{MC}(T_{eq})$ (black dash-dotted line in Figure~\ref{fig5} b)). Note that the pure PIC simulation (dashed blue curve) of the same physical and numerical parameters reveals a much improved energy conservation compared to the PIC-MC case, highlighting that this deterioration is indeed due to the coupling between the PIC and MC algorithms.

\section{Numerical heating minimization strategies}
\label{S:4}
Mitigation of the MC-induced numerical heating in PIC-MC simulations requires controlling the stochastic component of the current density associated with the MC collisions. In a thermal plasma, the collision current fluctuations are random and independent at each cell, and effectively act as a source of white noise in the system [Figure~\ref{fig6} (a)]. We have evaluated its effect on the emitted radiation by explicitly simulating the electromagnetic field produced by a stochastic current density distribution that mimics that due to MC collisions. We observe that the spectrum of the radiated electromagnetic energy is also broad, but is more pronounced at high spatial frequencies [Figure~\ref{fig6} (b)]. This suggests that applying a low-pass filter to the fields may be effective in mitigating the MC-induced heating. We have tested this strategy by using a compensated 5 pass binomial filter (applying a $1,2,1$ stencil $4$ times, followed by a $-5,14,-5$ stencil) on the electromagnetic fields every $\simeq 30$ collision periods ($\simeq500$ time steps) on the same simulation presented in Figure~\ref{fig1} (c). The results are presented in Figure~\ref{fig6} (c), and confirm that the heating rate can be effectively reduced ($\simeq25\times$ lower) by periodically filtering the electromagnetic fields. Note that the details of the filtering prescription will depend on the physics of interest in a particular simulation. Therefore, the optimal filtering strategy will need to be evaluated on a case by case basis.

Alternative to filtering techniques, improvements to the PIC-MC coupling may be possible by performing explicit corrections to the electromagnetic field or the current density on the grid, effectively subtracting the artificially produced radiation or the collisional current density and eliminating the heating. However, special care must be taken to ensure that such corrections do not deteriorate charge conservation.

\begin{figure}[t]
\begin{center}
\includegraphics[width=1.\linewidth]{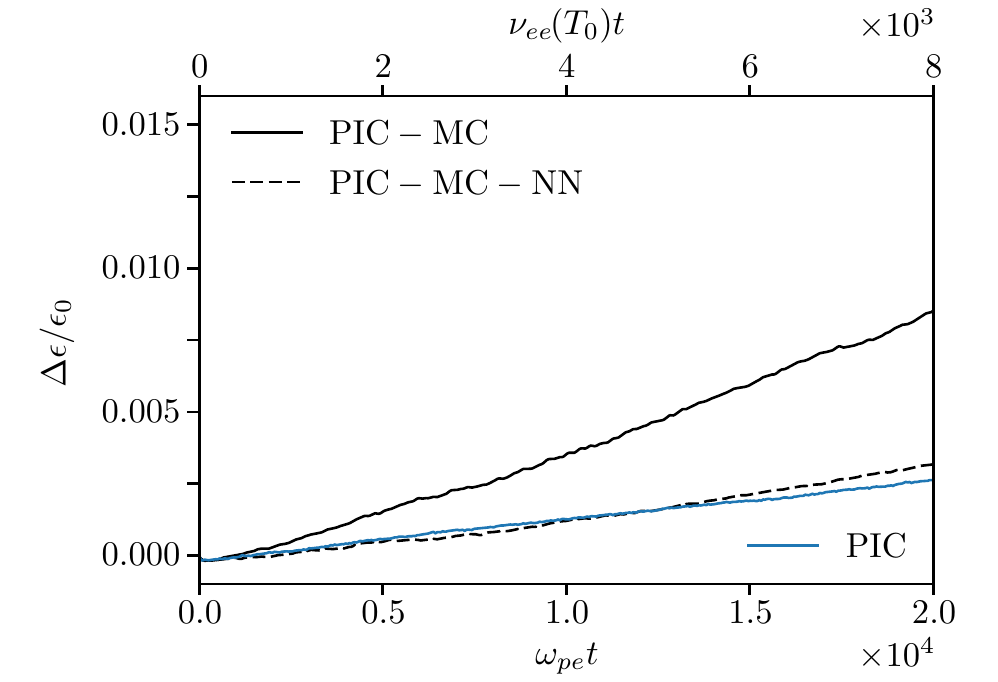}
\caption{Illustration of the effect of using different pair selection strategies for the MC collisions on the MC-induced numerical heating. In this example, we simulate a 1D spatially uniform electron-proton plasma in thermal equilibrium with the same physical and numerical parameters used in Figure~\ref{fig1}. However, in this case collisions were only performed between electrons. Electron-proton and proton-proton collisions were turned off. Shown are the deterioration of energy conservation for PIC-MC with the standard random pairing strategy (black solid curve), PIC-MC with nearest-neighbor (NN) pairing (black dashed curve) and for pure PIC (solid blue curve). The MC-induced heating is strongly reduced when using the NN pairing strategy, remaining only the innate PIC heating level. Note that the NN pairing strategy only reduces the MC-induced heating when collisions are performed between particles of the same charge to mass ratio.}
\label{fig7}
\end{center}
\end{figure}



It is interesting to note that for the special case of binary collisions between species of the same charge to mass ratio $q/m$, one can employ a different strategy to reduce the MC-induced heating rate. In this case, one sees from Eq.~\ref{eq:deltaj} that the collision current is proportional to the spatial separation between the collision pairs $x_{\alpha\beta} = x_{\alpha}-x_{\beta}$. For collision pairs that are randomly paired within a cell, the average spatial separation in 1D is $\langle x_{\alpha\beta} \rangle = \Delta x/3$. Yet, if one were to employ a different pairing strategy such as pairing nearest neighbor (NN) particles, the average spatial separation would be $\langle x_{\alpha\beta} \rangle = \Delta x/N_{ppc}$, which is $\ll \Delta x/3$ for commonly used numbers of particles per cell. In this case, the MC-induced heating using NN pairing becomes proportional to $1/N_{ppc}^3$, 
\begin{equation}
  \label{eq:gamma_mc3}
  \Gamma_\mathrm{MC-NN}(T_0) \propto \frac{ \nu_{\alpha\alpha}(T_0) (\omega_{pe}\Delta t)^2}{N_{ppc}^3},
\end{equation}

allowing to effectively suppress the numerical heating with a moderate number of particles per cell.

An illustration of this effect is shown in Figure~\ref{fig7}, where the MC-induced heating is measured in a 1D simulation of an electron-ion plasma where only electron-electron collisions are captured (electron-ion and ion-ion collisions are turned off.) This simulation used $100$ particles per cell, and one clearly observes that the MC-induced heating rate is strongly suppressed in this case when using the NN pairing strategy (black dashed curve). The remaining heating observed in the PIC-MC-NN case is that associated with the PIC method itself (blue solid curve).

The NN pairing strategy employed in our tests was achieved by sorting particles by their position within a cell in 1D. In 2 and 3-dimensional simulations, one can improve the locality of the collisions by using collision cells smaller than the PIC cell, and perform random pairing within these smaller collision cells. This would reduce the heating rate by the square of the ratio of the collision cell size to the PIC cell size.

Lastly, we note that it can be shown that low-energy (thermal) electrons can contribute significantly more to the MC-induced heating than high-energy (suprathermal) electrons in a thermal plasma. This is a consequence of low-energy electrons being more numerous and experiencing stronger collisional interactions than high-energy particles, and thus contribute more to $\delta\mathbf{J}_\mathrm{MC}$ on the grid. For instance, electrons with $v<2\sqrt{T/m_e}$ contribute $\sim 100\times$ more to the variance of $\delta\mathbf{J}_\mathrm{MC}$ (and hence to the MC-induced heating) than electrons with $v>2\sqrt{T/m_e}$. Thus, depending on the problem of interest, it may be viable to artificially lower the collision frequency of low-energy particles to mitigate the MC-induced heating without significantly impacting transport properties, which are governed by high-energy particles. These different strategies will be the focus of a follow up study.

\section{Conclusions}
\label{S:5}
We have shown that the coupling between the (energy-momentum conserving) binary MC collisions algorithm and the PIC algorithm leads to artificial heating of the system that is not present when these algorithms operate independently. We have shown that the numerical heating results from the inconsistency between the particle motion due to MC-collisions and the value of the collective electromagnetic fields on the PIC grid. The motion of the particles due to MC collisions results in the artificial production of electromagnetic radiation that heats the surrounding plasma particles. For typical numerical parameters used to model large-scale collisional systems ($\Delta t \simeq 0.2 \omega_{pe}^{-1}$, $N_{ppc} = 100$, $25$ and $8$ particles per cell in $1D$, $2D$,  and $3D$ simulations, and using high-order particle shapes to minimize numerical heating effects intrinsic to PIC itself), the MC-induced numerical heating can exceed $1\%$ after only $500$ collision periods ($1/\nu_{\alpha\beta}$).

Using a large enough number of particles per cell and small enough time step, it is possible to keep the heating rate at a tolerable level for a given simulation. However, if using more computational resources is unfeasible, we have shown it may be possible to reduce the MC-induced heating rate by periodically filtering the high-frequency electromagnetic fluctuations produced by the MC collisions. We also shown that for the special case of collisions between particles of equal charge-to-mass ratio, the collision-induced current can be effectively suppressed by modifying the collision pairing strategy. By selecting collision pairs that are closest to each other, the collision-induced current is reduced, and the associated numerical heating rate is significantly diminished.

While the work presented here focused on the coupling between binary MC collisions with PIC, we expect that the coupling between grid-based collision methods and PIC may suffer from a similar problem. In grid-based methods, a stochastic current density is also expected to be injected into the system, resulting in a rate proportional to $\Delta t ~\mathrm{Var}(\mathbf{\delta J}_\mathrm{MC})$. The derivation of the heating rate for grid-based methods should follow a similar procedure to the one outlined in this work.

The scalings obtained here for the MC-induced heating rate provide an important guide to determine suitable numerical parameters for large-scale PIC simulations of collisional plasmas. Future efforts will focus on the development of new strategies to minimize the variance of the stochastic current density that underlies this parasitic effect.

\section{Acknowledgements}
We thank the anonymous referee for an insightful suggestion regarding the mitigation of the radiation induced by low-energy particles. We thank Dr. V. K. Decyk and Dr. J. May for useful discussions. This work was supported by the U.S. Department of Energy (DOE) SLAC Contract No. DE-AC02-76SF00515, by the U.S. DOE Office of Science, Fusion Energy Sciences under Grants No. FWP 100237 and No. FWP 100182, and by the U.S. DOE Early Career Research Program under Grant No. FWP 100331. It was also partially supported by DOE Contracts No. DE-DE-SC0019010 and  DE-NA0003842, a UCOP award, LFR-17- 449059, and a U.S. NSF Grant No. ACI-1339893 (PICKSC). Simulations were run on Mira and Theta (ALCF) through ALCC awards.

\bibliography{scibib.bib}

\begin{thebibliography}{41}%
\makeatletter
\providecommand \@ifxundefined [1]{%
 \@ifx{#1\undefined}
}%
\providecommand \@ifnum [1]{%
 \ifnum #1\expandafter \@firstoftwo
 \else \expandafter \@secondoftwo
 \fi
}%
\providecommand \@ifx [1]{%
 \ifx #1\expandafter \@firstoftwo
 \else \expandafter \@secondoftwo
 \fi
}%
\providecommand \natexlab [1]{#1}%
\providecommand \enquote  [1]{``#1''}%
\providecommand \bibnamefont  [1]{#1}%
\providecommand \bibfnamefont [1]{#1}%
\providecommand \citenamefont [1]{#1}%
\providecommand \href@noop [0]{\@secondoftwo}%
\providecommand \href [0]{\begingroup \@sanitize@url \@href}%
\providecommand \@href[1]{\@@startlink{#1}\@@href}%
\providecommand \@@href[1]{\endgroup#1\@@endlink}%
\providecommand \@sanitize@url [0]{\catcode `\\12\catcode `\$12\catcode
  `\&12\catcode `\#12\catcode `\^12\catcode `\_12\catcode `\%12\relax}%
\providecommand \@@startlink[1]{}%
\providecommand \@@endlink[0]{}%
\providecommand \url  [0]{\begingroup\@sanitize@url \@url }%
\providecommand \@url [1]{\endgroup\@href {#1}{\urlprefix }}%
\providecommand \urlprefix  [0]{URL }%
\providecommand \Eprint [0]{\href }%
\providecommand \doibase [0]{http://dx.doi.org/}%
\providecommand \selectlanguage [0]{\@gobble}%
\providecommand \bibinfo  [0]{\@secondoftwo}%
\providecommand \bibfield  [0]{\@secondoftwo}%
\providecommand \translation [1]{[#1]}%
\providecommand \BibitemOpen [0]{}%
\providecommand \bibitemStop [0]{}%
\providecommand \bibitemNoStop [0]{.\EOS\space}%
\providecommand \EOS [0]{\spacefactor3000\relax}%
\providecommand \BibitemShut  [1]{\csname bibitem#1\endcsname}%
\let\auto@bib@innerbib\@empty
\bibitem [{\citenamefont {Dawson}(1983)}]{Dawson1983}%
  \BibitemOpen
  \bibfield  {author} {\bibinfo {author} {\bibfnamefont {J.~M.}\ \bibnamefont
  {Dawson}},\ }\href@noop {} {\bibfield  {journal} {\bibinfo  {journal}
  {Reviews of Modern Physics}\ }\textbf {\bibinfo {volume} {55}},\ \bibinfo
  {pages} {403} (\bibinfo {year} {1983})}\BibitemShut {NoStop}%
\bibitem [{\citenamefont {Hockney}\ and\ \citenamefont
  {Eastwood}(1988)}]{Hockney1988}%
  \BibitemOpen
  \bibfield  {author} {\bibinfo {author} {\bibfnamefont {R.~W.}\ \bibnamefont
  {Hockney}}\ and\ \bibinfo {author} {\bibfnamefont {J.~W.}\ \bibnamefont
  {Eastwood}},\ }\href@noop {} {\emph {\bibinfo {title} {Computer Simulation
  Using Particles}}}\ (\bibinfo  {publisher} {Taylor \& Francis, Inc.},\
  \bibinfo {address} {USA},\ \bibinfo {year} {1988})\BibitemShut {NoStop}%
\bibitem [{\citenamefont {Birdsall}\ and\ \citenamefont
  {Langdon}(1985)}]{Birdsall1985}%
  \BibitemOpen
  \bibfield  {author} {\bibinfo {author} {\bibfnamefont {C.~K.}\ \bibnamefont
  {Birdsall}}\ and\ \bibinfo {author} {\bibfnamefont {A.~B.}\ \bibnamefont
  {Langdon}},\ }\href@noop {} {\emph {\bibinfo {title} {Plasma Physics Via
  Computer}}}\ (\bibinfo  {publisher} {McGraw-Hill, Inc.},\ \bibinfo {address}
  {USA},\ \bibinfo {year} {1985})\BibitemShut {NoStop}%
\bibitem [{\citenamefont {Tajima}\ and\ \citenamefont
  {Dawson}(1979)}]{TajimaDawson1979}%
  \BibitemOpen
  \bibfield  {author} {\bibinfo {author} {\bibfnamefont {T.}~\bibnamefont
  {Tajima}}\ and\ \bibinfo {author} {\bibfnamefont {J.~M.}\ \bibnamefont
  {Dawson}},\ }\href@noop {} {\bibfield  {journal} {\bibinfo  {journal}
  {Physical Review Letters}\ }\textbf {\bibinfo {volume} {43}},\ \bibinfo
  {pages} {267} (\bibinfo {year} {1979})}\BibitemShut {NoStop}%
\bibitem [{\citenamefont {Litos}\ \emph {et~al.}(2014)\citenamefont {Litos},
  \citenamefont {Adli}, \citenamefont {An}, \citenamefont {Clarke},
  \citenamefont {Clayton}, \citenamefont {Corde}, \citenamefont {Delahaye},
  \citenamefont {England}, \citenamefont {Fisher}, \citenamefont {Frederico},
  \citenamefont {Gessner}, \citenamefont {Green}, \citenamefont {Hogan},
  \citenamefont {Joshi}, \citenamefont {Lu}, \citenamefont {Marsh},
  \citenamefont {Mori}, \citenamefont {Muggli}, \citenamefont
  {Vafaei-Najafabadi}, \citenamefont {Walz}, \citenamefont {White},
  \citenamefont {Wu}, \citenamefont {Yakimenko},\ and\ \citenamefont
  {Yocky}}]{Litos2014}%
  \BibitemOpen
  \bibfield  {author} {\bibinfo {author} {\bibfnamefont {M.}~\bibnamefont
  {Litos}}, \bibinfo {author} {\bibfnamefont {E.}~\bibnamefont {Adli}},
  \bibinfo {author} {\bibfnamefont {W.}~\bibnamefont {An}}, \bibinfo {author}
  {\bibfnamefont {C.~I.}\ \bibnamefont {Clarke}}, \bibinfo {author}
  {\bibfnamefont {C.~E.}\ \bibnamefont {Clayton}}, \bibinfo {author}
  {\bibfnamefont {S.}~\bibnamefont {Corde}}, \bibinfo {author} {\bibfnamefont
  {J.~P.}\ \bibnamefont {Delahaye}}, \bibinfo {author} {\bibfnamefont {R.~J.}\
  \bibnamefont {England}}, \bibinfo {author} {\bibfnamefont {A.~S.}\
  \bibnamefont {Fisher}}, \bibinfo {author} {\bibfnamefont {J.}~\bibnamefont
  {Frederico}}, \bibinfo {author} {\bibfnamefont {S.}~\bibnamefont {Gessner}},
  \bibinfo {author} {\bibfnamefont {S.~Z.}\ \bibnamefont {Green}}, \bibinfo
  {author} {\bibfnamefont {M.~J.}\ \bibnamefont {Hogan}}, \bibinfo {author}
  {\bibfnamefont {C.}~\bibnamefont {Joshi}}, \bibinfo {author} {\bibfnamefont
  {W.}~\bibnamefont {Lu}}, \bibinfo {author} {\bibfnamefont {K.~A.}\
  \bibnamefont {Marsh}}, \bibinfo {author} {\bibfnamefont {W.~B.}\ \bibnamefont
  {Mori}}, \bibinfo {author} {\bibfnamefont {P.}~\bibnamefont {Muggli}},
  \bibinfo {author} {\bibfnamefont {N.}~\bibnamefont {Vafaei-Najafabadi}},
  \bibinfo {author} {\bibfnamefont {D.}~\bibnamefont {Walz}}, \bibinfo {author}
  {\bibfnamefont {G.}~\bibnamefont {White}}, \bibinfo {author} {\bibfnamefont
  {Z.}~\bibnamefont {Wu}}, \bibinfo {author} {\bibfnamefont {V.}~\bibnamefont
  {Yakimenko}}, \ and\ \bibinfo {author} {\bibfnamefont {G.}~\bibnamefont
  {Yocky}},\ }\href@noop {} {\bibfield  {journal} {\bibinfo  {journal}
  {Nature}\ }\textbf {\bibinfo {volume} {515}},\ \bibinfo {pages} {92}
  (\bibinfo {year} {2014})}\BibitemShut {NoStop}%
\bibitem [{\citenamefont {Fi{\'u}za}\ \emph {et~al.}(2012)\citenamefont
  {Fi{\'u}za}, \citenamefont {Fonseca}, \citenamefont {Tonge}, \citenamefont
  {Mori},\ and\ \citenamefont {Silva}}]{Fiuza:2012kd}%
  \BibitemOpen
  \bibfield  {author} {\bibinfo {author} {\bibfnamefont {F.}~\bibnamefont
  {Fi{\'u}za}}, \bibinfo {author} {\bibfnamefont {R.~A.}\ \bibnamefont
  {Fonseca}}, \bibinfo {author} {\bibfnamefont {J.}~\bibnamefont {Tonge}},
  \bibinfo {author} {\bibfnamefont {W.~B.}\ \bibnamefont {Mori}}, \ and\
  \bibinfo {author} {\bibfnamefont {L.~O.}\ \bibnamefont {Silva}},\ }\href@noop
  {} {\bibfield  {journal} {\bibinfo  {journal} {Physical Review Letters}\
  }\textbf {\bibinfo {volume} {108}},\ \bibinfo {pages} {235004} (\bibinfo
  {year} {2012})}\BibitemShut {NoStop}%
\bibitem [{\citenamefont {Ridgers}\ \emph {et~al.}(2012)\citenamefont
  {Ridgers}, \citenamefont {Brady}, \citenamefont {Duclous}, \citenamefont
  {Kirk}, \citenamefont {Bennett}, \citenamefont {Arber}, \citenamefont
  {Robinson},\ and\ \citenamefont {Bell}}]{Ridgers2012}%
  \BibitemOpen
  \bibfield  {author} {\bibinfo {author} {\bibfnamefont {C.~P.}\ \bibnamefont
  {Ridgers}}, \bibinfo {author} {\bibfnamefont {C.~S.}\ \bibnamefont {Brady}},
  \bibinfo {author} {\bibfnamefont {R.}~\bibnamefont {Duclous}}, \bibinfo
  {author} {\bibfnamefont {J.~G.}\ \bibnamefont {Kirk}}, \bibinfo {author}
  {\bibfnamefont {K.}~\bibnamefont {Bennett}}, \bibinfo {author} {\bibfnamefont
  {T.~D.}\ \bibnamefont {Arber}}, \bibinfo {author} {\bibfnamefont {A.~P.~L.}\
  \bibnamefont {Robinson}}, \ and\ \bibinfo {author} {\bibfnamefont {A.~R.}\
  \bibnamefont {Bell}},\ }\href {\doibase 10.1103/PhysRevLett.108.165006}
  {\bibfield  {journal} {\bibinfo  {journal} {Physical Reviw Letters}\ }\textbf
  {\bibinfo {volume} {108}},\ \bibinfo {pages} {165006} (\bibinfo {year}
  {2012})}\BibitemShut {NoStop}%
\bibitem [{\citenamefont {Silva}\ \emph {et~al.}(2003)\citenamefont {Silva},
  \citenamefont {Fonseca}, \citenamefont {Tonge}, \citenamefont {Dawson},
  \citenamefont {Mori},\ and\ \citenamefont {Medvedev}}]{Silva:2003tq}%
  \BibitemOpen
  \bibfield  {author} {\bibinfo {author} {\bibfnamefont {L.~O.}\ \bibnamefont
  {Silva}}, \bibinfo {author} {\bibfnamefont {R.~A.}\ \bibnamefont {Fonseca}},
  \bibinfo {author} {\bibfnamefont {J.~W.}\ \bibnamefont {Tonge}}, \bibinfo
  {author} {\bibfnamefont {J.~M.}\ \bibnamefont {Dawson}}, \bibinfo {author}
  {\bibfnamefont {W.~B.}\ \bibnamefont {Mori}}, \ and\ \bibinfo {author}
  {\bibfnamefont {M.~V.}\ \bibnamefont {Medvedev}},\ }\href@noop {} {\bibfield
  {journal} {\bibinfo  {journal} {Astrophysical Journal}\ }\textbf {\bibinfo
  {volume} {596}},\ \bibinfo {pages} {L121} (\bibinfo {year}
  {2003})}\BibitemShut {NoStop}%
\bibitem [{\citenamefont {Alves}\ \emph {et~al.}(2012)\citenamefont {Alves},
  \citenamefont {Grismayer}, \citenamefont {Martins}, \citenamefont
  {Fi{\'u}za}, \citenamefont {Fonseca},\ and\ \citenamefont
  {Silva}}]{Alves:2012vb}%
  \BibitemOpen
  \bibfield  {author} {\bibinfo {author} {\bibfnamefont {E.~P.}\ \bibnamefont
  {Alves}}, \bibinfo {author} {\bibfnamefont {T.}~\bibnamefont {Grismayer}},
  \bibinfo {author} {\bibfnamefont {S.~F.}\ \bibnamefont {Martins}}, \bibinfo
  {author} {\bibfnamefont {F.}~\bibnamefont {Fi{\'u}za}}, \bibinfo {author}
  {\bibfnamefont {R.~A.}\ \bibnamefont {Fonseca}}, \ and\ \bibinfo {author}
  {\bibfnamefont {L.~O.}\ \bibnamefont {Silva}},\ }\href@noop {} {\bibfield
  {journal} {\bibinfo  {journal} {The Astrophysical Journal Letters}\ }\textbf
  {\bibinfo {volume} {746}},\ \bibinfo {pages} {L14} (\bibinfo {year}
  {2012})}\BibitemShut {NoStop}%
\bibitem [{\citenamefont {Huntington}\ \emph {et~al.}(2015)\citenamefont
  {Huntington}, \citenamefont {Fi{\'u}za}, \citenamefont {Ross}, \citenamefont
  {Zylstra}, \citenamefont {Drake}, \citenamefont {Froula}, \citenamefont
  {Gregori}, \citenamefont {Kugland}, \citenamefont {Kuranz}, \citenamefont
  {Levy}, \citenamefont {Li}, \citenamefont {Meinecke}, \citenamefont {Morita},
  \citenamefont {Petrasso}, \citenamefont {Plechaty}, \citenamefont
  {Remington}, \citenamefont {Ryutov}, \citenamefont {Sakawa}, \citenamefont
  {Spitkovsky}, \citenamefont {Takabe},\ and\ \citenamefont
  {Park}}]{Huntington:2015ko}%
  \BibitemOpen
  \bibfield  {author} {\bibinfo {author} {\bibfnamefont {C.~M.}\ \bibnamefont
  {Huntington}}, \bibinfo {author} {\bibfnamefont {F.}~\bibnamefont
  {Fi{\'u}za}}, \bibinfo {author} {\bibfnamefont {J.~S.}\ \bibnamefont {Ross}},
  \bibinfo {author} {\bibfnamefont {A.~B.}\ \bibnamefont {Zylstra}}, \bibinfo
  {author} {\bibfnamefont {R.~P.}\ \bibnamefont {Drake}}, \bibinfo {author}
  {\bibfnamefont {D.~H.}\ \bibnamefont {Froula}}, \bibinfo {author}
  {\bibfnamefont {G.}~\bibnamefont {Gregori}}, \bibinfo {author} {\bibfnamefont
  {N.~L.}\ \bibnamefont {Kugland}}, \bibinfo {author} {\bibfnamefont {C.~C.}\
  \bibnamefont {Kuranz}}, \bibinfo {author} {\bibfnamefont {M.~C.}\
  \bibnamefont {Levy}}, \bibinfo {author} {\bibfnamefont {C.~K.}\ \bibnamefont
  {Li}}, \bibinfo {author} {\bibfnamefont {J.}~\bibnamefont {Meinecke}},
  \bibinfo {author} {\bibfnamefont {T.}~\bibnamefont {Morita}}, \bibinfo
  {author} {\bibfnamefont {R.}~\bibnamefont {Petrasso}}, \bibinfo {author}
  {\bibfnamefont {C.}~\bibnamefont {Plechaty}}, \bibinfo {author}
  {\bibfnamefont {B.~A.}\ \bibnamefont {Remington}}, \bibinfo {author}
  {\bibfnamefont {D.~D.}\ \bibnamefont {Ryutov}}, \bibinfo {author}
  {\bibfnamefont {Y.}~\bibnamefont {Sakawa}}, \bibinfo {author} {\bibfnamefont
  {A.}~\bibnamefont {Spitkovsky}}, \bibinfo {author} {\bibfnamefont
  {H.}~\bibnamefont {Takabe}}, \ and\ \bibinfo {author} {\bibfnamefont {H.~S.}\
  \bibnamefont {Park}},\ }\href@noop {} {\bibfield  {journal} {\bibinfo
  {journal} {Nature Physics}\ }\textbf {\bibinfo {volume} {11}},\ \bibinfo
  {pages} {173} (\bibinfo {year} {2015})}\BibitemShut {NoStop}%
\bibitem [{\citenamefont {Alves}\ \emph {et~al.}(2019)\citenamefont {Alves},
  \citenamefont {Zrake},\ and\ \citenamefont {Fi{\'u}za}}]{Alves2019}%
  \BibitemOpen
  \bibfield  {author} {\bibinfo {author} {\bibfnamefont {E.~P.}\ \bibnamefont
  {Alves}}, \bibinfo {author} {\bibfnamefont {J.}~\bibnamefont {Zrake}}, \ and\
  \bibinfo {author} {\bibfnamefont {F.}~\bibnamefont {Fi{\'u}za}},\ }\href@noop
  {} {\bibfield  {journal} {\bibinfo  {journal} {Physics of Plasmas}\ }\textbf
  {\bibinfo {volume} {26}},\ \bibinfo {pages} {072105} (\bibinfo {year}
  {2019})}\BibitemShut {NoStop}%
\bibitem [{\citenamefont {Alves}\ \emph {et~al.}(2018)\citenamefont {Alves},
  \citenamefont {Zrake},\ and\ \citenamefont {Fi{\'u}za}}]{Alves2018}%
  \BibitemOpen
  \bibfield  {author} {\bibinfo {author} {\bibfnamefont {E.~P.}\ \bibnamefont
  {Alves}}, \bibinfo {author} {\bibfnamefont {J.}~\bibnamefont {Zrake}}, \ and\
  \bibinfo {author} {\bibfnamefont {F.}~\bibnamefont {Fi{\'u}za}},\ }\href@noop
  {} {\bibfield  {journal} {\bibinfo  {journal} {Physical Review Letters}\
  }\textbf {\bibinfo {volume} {121}},\ \bibinfo {pages} {245101} (\bibinfo
  {year} {2018})}\BibitemShut {NoStop}%
\bibitem [{\citenamefont {Spitkovsky}(2008)}]{Spitkovsky:2008wo}%
  \BibitemOpen
  \bibfield  {author} {\bibinfo {author} {\bibfnamefont {A.}~\bibnamefont
  {Spitkovsky}},\ }\href@noop {} {\bibfield  {journal} {\bibinfo  {journal}
  {Astrophysical Journal Letters}\ }\textbf {\bibinfo {volume} {682}},\
  \bibinfo {pages} {L5} (\bibinfo {year} {2008})}\BibitemShut {NoStop}%
\bibitem [{\citenamefont {Guo}\ \emph {et~al.}(2014)\citenamefont {Guo},
  \citenamefont {Li}, \citenamefont {Daughton},\ and\ \citenamefont
  {Liu}}]{Guo:2014gs}%
  \BibitemOpen
  \bibfield  {author} {\bibinfo {author} {\bibfnamefont {F.}~\bibnamefont
  {Guo}}, \bibinfo {author} {\bibfnamefont {H.}~\bibnamefont {Li}}, \bibinfo
  {author} {\bibfnamefont {W.}~\bibnamefont {Daughton}}, \ and\ \bibinfo
  {author} {\bibfnamefont {Y.-H.}\ \bibnamefont {Liu}},\ }\href@noop {}
  {\bibfield  {journal} {\bibinfo  {journal} {Physical Review Letters}\
  }\textbf {\bibinfo {volume} {113}},\ \bibinfo {pages} {155005} (\bibinfo
  {year} {2014})}\BibitemShut {NoStop}%
\bibitem [{not()}]{note1}%
  \BibitemOpen
  \href@noop {} {}\bibinfo {note} {Note that if the PIC grid resolution were to
  resolve the classical electron radius, then the self-consistent
  inter-particle electromagnetic field would be correctly calculated, but this
  is not done in practice.}\BibitemShut {Stop}%
\bibitem [{\citenamefont {Lemons}\ \emph {et~al.}(2009)\citenamefont {Lemons},
  \citenamefont {Winske}, \citenamefont {Daughton},\ and\ \citenamefont
  {Albright}}]{Lemons:2009da}%
  \BibitemOpen
  \bibfield  {author} {\bibinfo {author} {\bibfnamefont {D.~S.}\ \bibnamefont
  {Lemons}}, \bibinfo {author} {\bibfnamefont {D.}~\bibnamefont {Winske}},
  \bibinfo {author} {\bibfnamefont {W.}~\bibnamefont {Daughton}}, \ and\
  \bibinfo {author} {\bibfnamefont {B.}~\bibnamefont {Albright}},\ }\href@noop
  {} {\bibfield  {journal} {\bibinfo  {journal} {Journal of Computational
  Physics}\ }\textbf {\bibinfo {volume} {228}},\ \bibinfo {pages} {1391}
  (\bibinfo {year} {2009})}\BibitemShut {NoStop}%
\bibitem [{\citenamefont {Cohen}\ \emph {et~al.}(2013)\citenamefont {Cohen},
  \citenamefont {Dimits},\ and\ \citenamefont {Strozzi}}]{Cohen:2013if}%
  \BibitemOpen
  \bibfield  {author} {\bibinfo {author} {\bibfnamefont {B.~I.}\ \bibnamefont
  {Cohen}}, \bibinfo {author} {\bibfnamefont {A.~M.}\ \bibnamefont {Dimits}}, \
  and\ \bibinfo {author} {\bibfnamefont {D.~J.}\ \bibnamefont {Strozzi}},\
  }\href@noop {} {\bibfield  {journal} {\bibinfo  {journal} {Journal of
  Computational Physics}\ }\textbf {\bibinfo {volume} {234}},\ \bibinfo {pages}
  {33} (\bibinfo {year} {2013})}\BibitemShut {NoStop}%
\bibitem [{\citenamefont {Takizuka}\ and\ \citenamefont
  {Abe}(1977)}]{TakizukaAbe1977}%
  \BibitemOpen
  \bibfield  {author} {\bibinfo {author} {\bibfnamefont {T.}~\bibnamefont
  {Takizuka}}\ and\ \bibinfo {author} {\bibfnamefont {H.}~\bibnamefont {Abe}},\
  }\href@noop {} {\bibfield  {journal} {\bibinfo  {journal} {Journal of
  Computational Physics}\ }\textbf {\bibinfo {volume} {25}},\ \bibinfo {pages}
  {205} (\bibinfo {year} {1977})}\BibitemShut {NoStop}%
\bibitem [{\citenamefont {Nanbu}(1997)}]{Nanbu1997}%
  \BibitemOpen
  \bibfield  {author} {\bibinfo {author} {\bibfnamefont {K.}~\bibnamefont
  {Nanbu}},\ }\href@noop {} {\bibfield  {journal} {\bibinfo  {journal}
  {Physical Review E}\ }\textbf {\bibinfo {volume} {55}},\ \bibinfo {pages}
  {4642} (\bibinfo {year} {1997})}\BibitemShut {NoStop}%
\bibitem [{\citenamefont {Nanbu}\ and\ \citenamefont
  {Yonemura}(1998)}]{Nanbu1998}%
  \BibitemOpen
  \bibfield  {author} {\bibinfo {author} {\bibfnamefont {K.}~\bibnamefont
  {Nanbu}}\ and\ \bibinfo {author} {\bibfnamefont {S.}~\bibnamefont
  {Yonemura}},\ }\href@noop {} {\bibfield  {journal} {\bibinfo  {journal}
  {Journal of Computational Physics}\ }\textbf {\bibinfo {volume} {145}},\
  \bibinfo {pages} {639} (\bibinfo {year} {1998})}\BibitemShut {NoStop}%
\bibitem [{\citenamefont {Sentoku}\ and\ \citenamefont
  {Kemp}(2008)}]{Sentoku:2008iu}%
  \BibitemOpen
  \bibfield  {author} {\bibinfo {author} {\bibfnamefont {Y.}~\bibnamefont
  {Sentoku}}\ and\ \bibinfo {author} {\bibfnamefont {A.~J.}\ \bibnamefont
  {Kemp}},\ }\href@noop {} {\bibfield  {journal} {\bibinfo  {journal} {Journal
  of Computational Physics}\ }\textbf {\bibinfo {volume} {227}},\ \bibinfo
  {pages} {6846} (\bibinfo {year} {2008})}\BibitemShut {NoStop}%
\bibitem [{\citenamefont {P{\'e}rez}\ \emph {et~al.}(2012)\citenamefont
  {P{\'e}rez}, \citenamefont {Gremillet}, \citenamefont {Decoster},
  \citenamefont {Drouin},\ and\ \citenamefont {Lefebvre}}]{Perez:2012fy}%
  \BibitemOpen
  \bibfield  {author} {\bibinfo {author} {\bibfnamefont {F.}~\bibnamefont
  {P{\'e}rez}}, \bibinfo {author} {\bibfnamefont {L.}~\bibnamefont
  {Gremillet}}, \bibinfo {author} {\bibfnamefont {A.}~\bibnamefont {Decoster}},
  \bibinfo {author} {\bibfnamefont {M.}~\bibnamefont {Drouin}}, \ and\ \bibinfo
  {author} {\bibfnamefont {E.}~\bibnamefont {Lefebvre}},\ }\href@noop {}
  {\bibfield  {journal} {\bibinfo  {journal} {Physics of Plasmas}\ }\textbf
  {\bibinfo {volume} {19}},\ \bibinfo {pages} {083104} (\bibinfo {year}
  {2012})}\BibitemShut {NoStop}%
\bibitem [{\citenamefont {Fonseca}\ \emph {et~al.}(2002)\citenamefont
  {Fonseca}, \citenamefont {Silva}, \citenamefont {Tsung}, \citenamefont
  {Decyk}, \citenamefont {Lu}, \citenamefont {Ren}, \citenamefont {Mori},
  \citenamefont {Deng}, \citenamefont {Lee}, \citenamefont {Katsouleas},\ and\
  \citenamefont {Adam}}]{Fonseca2002}%
  \BibitemOpen
  \bibfield  {author} {\bibinfo {author} {\bibfnamefont {R.~A.}\ \bibnamefont
  {Fonseca}}, \bibinfo {author} {\bibfnamefont {L.~O.}\ \bibnamefont {Silva}},
  \bibinfo {author} {\bibfnamefont {F.~S.}\ \bibnamefont {Tsung}}, \bibinfo
  {author} {\bibfnamefont {V.~K.}\ \bibnamefont {Decyk}}, \bibinfo {author}
  {\bibfnamefont {W.}~\bibnamefont {Lu}}, \bibinfo {author} {\bibfnamefont
  {C.}~\bibnamefont {Ren}}, \bibinfo {author} {\bibfnamefont {W.~B.}\
  \bibnamefont {Mori}}, \bibinfo {author} {\bibfnamefont {S.}~\bibnamefont
  {Deng}}, \bibinfo {author} {\bibfnamefont {S.}~\bibnamefont {Lee}}, \bibinfo
  {author} {\bibfnamefont {T.}~\bibnamefont {Katsouleas}}, \ and\ \bibinfo
  {author} {\bibfnamefont {J.~C.}\ \bibnamefont {Adam}},\ }\href@noop {}
  {\bibfield  {journal} {\bibinfo  {journal} {Computational Science-ICCS 2002,
  Pt III, Proceedings}\ }\textbf {\bibinfo {volume} {2331}},\ \bibinfo {pages}
  {342} (\bibinfo {year} {2002})}\BibitemShut {NoStop}%
\bibitem [{\citenamefont {Fonseca}\ \emph {et~al.}(2008)\citenamefont
  {Fonseca}, \citenamefont {Martins}, \citenamefont {Silva}, \citenamefont
  {Tonge}, \citenamefont {Tsung},\ and\ \citenamefont {Mori}}]{Fonseca:2008ib}%
  \BibitemOpen
  \bibfield  {author} {\bibinfo {author} {\bibfnamefont {R.~A.}\ \bibnamefont
  {Fonseca}}, \bibinfo {author} {\bibfnamefont {S.~F.}\ \bibnamefont
  {Martins}}, \bibinfo {author} {\bibfnamefont {L.~O.}\ \bibnamefont {Silva}},
  \bibinfo {author} {\bibfnamefont {J.~W.}\ \bibnamefont {Tonge}}, \bibinfo
  {author} {\bibfnamefont {F.~S.}\ \bibnamefont {Tsung}}, \ and\ \bibinfo
  {author} {\bibfnamefont {W.~B.}\ \bibnamefont {Mori}},\ }\href@noop {}
  {\bibfield  {journal} {\bibinfo  {journal} {Plasma Physics and Controlled
  Fusion}\ }\textbf {\bibinfo {volume} {50}},\ \bibinfo {pages} {124034}
  (\bibinfo {year} {2008})}\BibitemShut {NoStop}%
\bibitem [{\citenamefont {Bowers}\ \emph {et~al.}(2008)\citenamefont {Bowers},
  \citenamefont {Albright}, \citenamefont {Yin}, \citenamefont {Bergen},\ and\
  \citenamefont {Kwan}}]{vpic}%
  \BibitemOpen
  \bibfield  {author} {\bibinfo {author} {\bibfnamefont {K.~J.}\ \bibnamefont
  {Bowers}}, \bibinfo {author} {\bibfnamefont {B.~J.}\ \bibnamefont
  {Albright}}, \bibinfo {author} {\bibfnamefont {L.}~\bibnamefont {Yin}},
  \bibinfo {author} {\bibfnamefont {B.}~\bibnamefont {Bergen}}, \ and\ \bibinfo
  {author} {\bibfnamefont {T.~J.~T.}\ \bibnamefont {Kwan}},\ }\href {\doibase
  10.1063/1.2840133} {\bibfield  {journal} {\bibinfo  {journal} {Physics of
  Plasmas}\ }\textbf {\bibinfo {volume} {15}},\ \bibinfo {pages} {055703}
  (\bibinfo {year} {2008})}\BibitemShut {NoStop}%
\bibitem [{\citenamefont {Lefebvre}\ \emph {et~al.}(2003)\citenamefont
  {Lefebvre}, \citenamefont {Cochet}, \citenamefont {Fritzler}, \citenamefont
  {Malka}, \citenamefont {onard}, \citenamefont {Chemin}, \citenamefont
  {Darbon}, \citenamefont {Disdier}, \citenamefont {Faure}, \citenamefont
  {Fedotoff}, \citenamefont {Landoas}, \citenamefont {Malka}, \citenamefont
  {ot}, \citenamefont {Morel}, \citenamefont {Gloahec}, \citenamefont {Rouyer},
  \citenamefont {Rubbelynck}, \citenamefont {Tikhonchuk}, \citenamefont
  {Wrobel}, \citenamefont {Audebert},\ and\ \citenamefont
  {Rousseaux}}]{calder}%
  \BibitemOpen
  \bibfield  {author} {\bibinfo {author} {\bibfnamefont {E.}~\bibnamefont
  {Lefebvre}}, \bibinfo {author} {\bibfnamefont {N.}~\bibnamefont {Cochet}},
  \bibinfo {author} {\bibfnamefont {S.}~\bibnamefont {Fritzler}}, \bibinfo
  {author} {\bibfnamefont {V.}~\bibnamefont {Malka}}, \bibinfo {author}
  {\bibfnamefont {M.-M.~A.}\ \bibnamefont {onard}}, \bibinfo {author}
  {\bibfnamefont {J.-F.}\ \bibnamefont {Chemin}}, \bibinfo {author}
  {\bibfnamefont {S.}~\bibnamefont {Darbon}}, \bibinfo {author} {\bibfnamefont
  {L.}~\bibnamefont {Disdier}}, \bibinfo {author} {\bibfnamefont
  {J.}~\bibnamefont {Faure}}, \bibinfo {author} {\bibfnamefont
  {A.}~\bibnamefont {Fedotoff}}, \bibinfo {author} {\bibfnamefont
  {O.}~\bibnamefont {Landoas}}, \bibinfo {author} {\bibfnamefont
  {G.}~\bibnamefont {Malka}}, \bibinfo {author} {\bibfnamefont {V.~M.}\
  \bibnamefont {ot}}, \bibinfo {author} {\bibfnamefont {P.}~\bibnamefont
  {Morel}}, \bibinfo {author} {\bibfnamefont {M.~R.~L.}\ \bibnamefont
  {Gloahec}}, \bibinfo {author} {\bibfnamefont {A.}~\bibnamefont {Rouyer}},
  \bibinfo {author} {\bibfnamefont {C.}~\bibnamefont {Rubbelynck}}, \bibinfo
  {author} {\bibfnamefont {V.}~\bibnamefont {Tikhonchuk}}, \bibinfo {author}
  {\bibfnamefont {R.}~\bibnamefont {Wrobel}}, \bibinfo {author} {\bibfnamefont
  {P.}~\bibnamefont {Audebert}}, \ and\ \bibinfo {author} {\bibfnamefont
  {C.}~\bibnamefont {Rousseaux}},\ }\href {\doibase 10.1088/0029-5515/43/7/317}
  {\bibfield  {journal} {\bibinfo  {journal} {Nuclear Fusion}\ }\textbf
  {\bibinfo {volume} {43}},\ \bibinfo {pages} {629} (\bibinfo {year}
  {2003})}\BibitemShut {NoStop}%
\bibitem [{\citenamefont {Derouillat}\ \emph {et~al.}(2018)\citenamefont
  {Derouillat}, \citenamefont {Beck}, \citenamefont {Pérez}, \citenamefont
  {Vinci}, \citenamefont {Chiaramello}, \citenamefont {Grassi}, \citenamefont
  {Flé}, \citenamefont {Bouchard}, \citenamefont {Plotnikov}, \citenamefont
  {Aunai}, \citenamefont {Dargent}, \citenamefont {Riconda},\ and\
  \citenamefont {Grech}}]{smilei}%
  \BibitemOpen
  \bibfield  {author} {\bibinfo {author} {\bibfnamefont {J.}~\bibnamefont
  {Derouillat}}, \bibinfo {author} {\bibfnamefont {A.}~\bibnamefont {Beck}},
  \bibinfo {author} {\bibfnamefont {F.}~\bibnamefont {Pérez}}, \bibinfo
  {author} {\bibfnamefont {T.}~\bibnamefont {Vinci}}, \bibinfo {author}
  {\bibfnamefont {M.}~\bibnamefont {Chiaramello}}, \bibinfo {author}
  {\bibfnamefont {A.}~\bibnamefont {Grassi}}, \bibinfo {author} {\bibfnamefont
  {M.}~\bibnamefont {Flé}}, \bibinfo {author} {\bibfnamefont {G.}~\bibnamefont
  {Bouchard}}, \bibinfo {author} {\bibfnamefont {I.}~\bibnamefont {Plotnikov}},
  \bibinfo {author} {\bibfnamefont {N.}~\bibnamefont {Aunai}}, \bibinfo
  {author} {\bibfnamefont {J.}~\bibnamefont {Dargent}}, \bibinfo {author}
  {\bibfnamefont {C.}~\bibnamefont {Riconda}}, \ and\ \bibinfo {author}
  {\bibfnamefont {M.}~\bibnamefont {Grech}},\ }\href {\doibase
  https://doi.org/10.1016/j.cpc.2017.09.024} {\bibfield  {journal} {\bibinfo
  {journal} {Computer Physics Communications}\ }\textbf {\bibinfo {volume}
  {222}},\ \bibinfo {pages} {351 } (\bibinfo {year} {2018})}\BibitemShut
  {NoStop}%
\bibitem [{\citenamefont {Arber}\ \emph {et~al.}(2015)\citenamefont {Arber},
  \citenamefont {Bennett}, \citenamefont {Brady}, \citenamefont
  {Lawrence-Douglas}, \citenamefont {Ramsay}, \citenamefont {Sircombe},
  \citenamefont {Gillies}, \citenamefont {Evans}, \citenamefont {Schmitz},
  \citenamefont {Bell},\ and\ \citenamefont {Ridgers}}]{epoch}%
  \BibitemOpen
  \bibfield  {author} {\bibinfo {author} {\bibfnamefont {T.~D.}\ \bibnamefont
  {Arber}}, \bibinfo {author} {\bibfnamefont {K.}~\bibnamefont {Bennett}},
  \bibinfo {author} {\bibfnamefont {C.~S.}\ \bibnamefont {Brady}}, \bibinfo
  {author} {\bibfnamefont {A.}~\bibnamefont {Lawrence-Douglas}}, \bibinfo
  {author} {\bibfnamefont {M.~G.}\ \bibnamefont {Ramsay}}, \bibinfo {author}
  {\bibfnamefont {N.~J.}\ \bibnamefont {Sircombe}}, \bibinfo {author}
  {\bibfnamefont {P.}~\bibnamefont {Gillies}}, \bibinfo {author} {\bibfnamefont
  {R.~G.}\ \bibnamefont {Evans}}, \bibinfo {author} {\bibfnamefont
  {H.}~\bibnamefont {Schmitz}}, \bibinfo {author} {\bibfnamefont {A.~R.}\
  \bibnamefont {Bell}}, \ and\ \bibinfo {author} {\bibfnamefont {C.~P.}\
  \bibnamefont {Ridgers}},\ }\href {\doibase 10.1088/0741-3335/57/11/113001}
  {\bibfield  {journal} {\bibinfo  {journal} {Plasma Physics and Controlled
  Fusion}\ }\textbf {\bibinfo {volume} {57}},\ \bibinfo {pages} {113001}
  (\bibinfo {year} {2015})}\BibitemShut {NoStop}%
\bibitem [{\citenamefont {Wang}\ \emph {et~al.}(2008)\citenamefont {Wang},
  \citenamefont {Lin}, \citenamefont {Caflisch}, \citenamefont {Cohen},\ and\
  \citenamefont {Dimits}}]{Wang:2008co}%
  \BibitemOpen
  \bibfield  {author} {\bibinfo {author} {\bibfnamefont {C.}~\bibnamefont
  {Wang}}, \bibinfo {author} {\bibfnamefont {T.}~\bibnamefont {Lin}}, \bibinfo
  {author} {\bibfnamefont {R.}~\bibnamefont {Caflisch}}, \bibinfo {author}
  {\bibfnamefont {B.~I.}\ \bibnamefont {Cohen}}, \ and\ \bibinfo {author}
  {\bibfnamefont {A.~M.}\ \bibnamefont {Dimits}},\ }\href@noop {} {\bibfield
  {journal} {\bibinfo  {journal} {Journal of Computational Physics}\ }\textbf
  {\bibinfo {volume} {227}},\ \bibinfo {pages} {4308} (\bibinfo {year}
  {2008})}\BibitemShut {NoStop}%
\bibitem [{\citenamefont {Cohen}\ \emph {et~al.}(2010)\citenamefont {Cohen},
  \citenamefont {Dimits}, \citenamefont {Friedman},\ and\ \citenamefont
  {Caflisch}}]{Cohen:cr}%
  \BibitemOpen
  \bibfield  {author} {\bibinfo {author} {\bibfnamefont {B.~I.}\ \bibnamefont
  {Cohen}}, \bibinfo {author} {\bibfnamefont {A.~M.}\ \bibnamefont {Dimits}},
  \bibinfo {author} {\bibfnamefont {A.}~\bibnamefont {Friedman}}, \ and\
  \bibinfo {author} {\bibfnamefont {R.~E.}\ \bibnamefont {Caflisch}},\
  }\href@noop {} {\bibfield  {journal} {\bibinfo  {journal} {Plasma Science,
  IEEE Transactions on}\ }\textbf {\bibinfo {volume} {38}},\ \bibinfo {pages}
  {2394} (\bibinfo {year} {2010})}\BibitemShut {NoStop}%
\bibitem [{\citenamefont {Peano}\ \emph {et~al.}(2009)\citenamefont {Peano},
  \citenamefont {Marti}, \citenamefont {Silva},\ and\ \citenamefont
  {Coppa}}]{Peano:2009iy}%
  \BibitemOpen
  \bibfield  {author} {\bibinfo {author} {\bibfnamefont {F.}~\bibnamefont
  {Peano}}, \bibinfo {author} {\bibfnamefont {M.}~\bibnamefont {Marti}},
  \bibinfo {author} {\bibfnamefont {L.~O.}\ \bibnamefont {Silva}}, \ and\
  \bibinfo {author} {\bibfnamefont {G.}~\bibnamefont {Coppa}},\ }\href
  {\doibase 10.1103/PhysRevE.79.025701} {\bibfield  {journal} {\bibinfo
  {journal} {Phys. Rev. E}\ }\textbf {\bibinfo {volume} {79}},\ \bibinfo
  {pages} {025701(R)} (\bibinfo {year} {2009})}\BibitemShut {NoStop}%
\bibitem [{\citenamefont {Higginson}(2017)}]{Higginson2017}%
  \BibitemOpen
  \bibfield  {author} {\bibinfo {author} {\bibfnamefont {D.~P.}\ \bibnamefont
  {Higginson}},\ }\href@noop {} {\bibfield  {journal} {\bibinfo  {journal}
  {Journal of Computational Physics}\ }\textbf {\bibinfo {volume} {349}},\
  \bibinfo {pages} {589} (\bibinfo {year} {2017})}\BibitemShut {NoStop}%
\bibitem [{\citenamefont {Amendt}\ \emph {et~al.}(2011)\citenamefont {Amendt},
  \citenamefont {Wilks}, \citenamefont {Bellei}, \citenamefont {Li},\ and\
  \citenamefont {Petrasso}}]{Amendt2011}%
  \BibitemOpen
  \bibfield  {author} {\bibinfo {author} {\bibfnamefont {P.}~\bibnamefont
  {Amendt}}, \bibinfo {author} {\bibfnamefont {S.~C.}\ \bibnamefont {Wilks}},
  \bibinfo {author} {\bibfnamefont {C.}~\bibnamefont {Bellei}}, \bibinfo
  {author} {\bibfnamefont {C.~K.}\ \bibnamefont {Li}}, \ and\ \bibinfo {author}
  {\bibfnamefont {R.~D.}\ \bibnamefont {Petrasso}},\ }\href@noop {} {\bibfield
  {journal} {\bibinfo  {journal} {Physics of Plasmas}\ }\textbf {\bibinfo
  {volume} {18}},\ \bibinfo {pages} {056308} (\bibinfo {year}
  {2011})}\BibitemShut {NoStop}%
\bibitem [{\citenamefont {Bellei}\ \emph {et~al.}(2014)\citenamefont {Bellei},
  \citenamefont {Rinderknecht}, \citenamefont {Zylstra}, \citenamefont
  {Rosenberg}, \citenamefont {Sio}, \citenamefont {Li}, \citenamefont
  {Petrasso}, \citenamefont {Wilks},\ and\ \citenamefont
  {Amendt}}]{Bellei:2014kq}%
  \BibitemOpen
  \bibfield  {author} {\bibinfo {author} {\bibfnamefont {C.}~\bibnamefont
  {Bellei}}, \bibinfo {author} {\bibfnamefont {H.}~\bibnamefont
  {Rinderknecht}}, \bibinfo {author} {\bibfnamefont {A.}~\bibnamefont
  {Zylstra}}, \bibinfo {author} {\bibfnamefont {M.}~\bibnamefont {Rosenberg}},
  \bibinfo {author} {\bibfnamefont {H.}~\bibnamefont {Sio}}, \bibinfo {author}
  {\bibfnamefont {C.~K.}\ \bibnamefont {Li}}, \bibinfo {author} {\bibfnamefont
  {R.}~\bibnamefont {Petrasso}}, \bibinfo {author} {\bibfnamefont {S.~C.}\
  \bibnamefont {Wilks}}, \ and\ \bibinfo {author} {\bibfnamefont {P.~A.}\
  \bibnamefont {Amendt}},\ }\href@noop {} {\bibfield  {journal} {\bibinfo
  {journal} {Physics of Plasmas}\ }\textbf {\bibinfo {volume} {21}},\ \bibinfo
  {pages} {056310} (\bibinfo {year} {2014})}\BibitemShut {NoStop}%
\bibitem [{\citenamefont {Bellei}\ and\ \citenamefont
  {Amendt}(2014)}]{Bellei:2014ks}%
  \BibitemOpen
  \bibfield  {author} {\bibinfo {author} {\bibfnamefont {C.}~\bibnamefont
  {Bellei}}\ and\ \bibinfo {author} {\bibfnamefont {P.~A.}\ \bibnamefont
  {Amendt}},\ }\href@noop {} {\bibfield  {journal} {\bibinfo  {journal}
  {Physical Review E}\ }\textbf {\bibinfo {volume} {90}},\ \bibinfo {pages}
  {013101} (\bibinfo {year} {2014})}\BibitemShut {NoStop}%
\bibitem [{\citenamefont {Rinderknecht}\ \emph {et~al.}(2015)\citenamefont
  {Rinderknecht}, \citenamefont {Rosenberg}, \citenamefont {Li}, \citenamefont
  {Hoffman}, \citenamefont {Kagan}, \citenamefont {Zylstra}, \citenamefont
  {Sio}, \citenamefont {Frenje}, \citenamefont {Gatu~Johnson}, \citenamefont
  {S{\'e}guin}, \citenamefont {Petrasso}, \citenamefont {Amendt}, \citenamefont
  {Bellei}, \citenamefont {Wilks}, \citenamefont {Delettrez}, \citenamefont
  {Glebov}, \citenamefont {Stoeckl}, \citenamefont {Sangster}, \citenamefont
  {Meyerhofer},\ and\ \citenamefont {Nikroo}}]{Rinderknecht:2015et}%
  \BibitemOpen
  \bibfield  {author} {\bibinfo {author} {\bibfnamefont {H.~G.}\ \bibnamefont
  {Rinderknecht}}, \bibinfo {author} {\bibfnamefont {M.~J.}\ \bibnamefont
  {Rosenberg}}, \bibinfo {author} {\bibfnamefont {C.~K.}\ \bibnamefont {Li}},
  \bibinfo {author} {\bibfnamefont {N.~M.}\ \bibnamefont {Hoffman}}, \bibinfo
  {author} {\bibfnamefont {G.}~\bibnamefont {Kagan}}, \bibinfo {author}
  {\bibfnamefont {A.~B.}\ \bibnamefont {Zylstra}}, \bibinfo {author}
  {\bibfnamefont {H.}~\bibnamefont {Sio}}, \bibinfo {author} {\bibfnamefont
  {J.~A.}\ \bibnamefont {Frenje}}, \bibinfo {author} {\bibfnamefont
  {M.}~\bibnamefont {Gatu~Johnson}}, \bibinfo {author} {\bibfnamefont {F.~H.}\
  \bibnamefont {S{\'e}guin}}, \bibinfo {author} {\bibfnamefont {R.~D.}\
  \bibnamefont {Petrasso}}, \bibinfo {author} {\bibfnamefont {P.}~\bibnamefont
  {Amendt}}, \bibinfo {author} {\bibfnamefont {C.}~\bibnamefont {Bellei}},
  \bibinfo {author} {\bibfnamefont {S.}~\bibnamefont {Wilks}}, \bibinfo
  {author} {\bibfnamefont {J.}~\bibnamefont {Delettrez}}, \bibinfo {author}
  {\bibfnamefont {V.~Y.}\ \bibnamefont {Glebov}}, \bibinfo {author}
  {\bibfnamefont {C.}~\bibnamefont {Stoeckl}}, \bibinfo {author} {\bibfnamefont
  {T.~C.}\ \bibnamefont {Sangster}}, \bibinfo {author} {\bibfnamefont {D.~D.}\
  \bibnamefont {Meyerhofer}}, \ and\ \bibinfo {author} {\bibfnamefont
  {A.}~\bibnamefont {Nikroo}},\ }\href@noop {} {\bibfield  {journal} {\bibinfo
  {journal} {Physical Review Letters}\ }\textbf {\bibinfo {volume} {114}},\
  \bibinfo {pages} {025001} (\bibinfo {year} {2015})}\BibitemShut {NoStop}%
\bibitem [{\citenamefont {Okuda}(1970)}]{Okuda1970}%
  \BibitemOpen
  \bibfield  {author} {\bibinfo {author} {\bibfnamefont {H.}~\bibnamefont
  {Okuda}},\ }\href@noop {} {\bibfield  {journal} {\bibinfo  {journal} {Physics
  of Fluids}\ }\textbf {\bibinfo {volume} {13}},\ \bibinfo {pages} {2123}
  (\bibinfo {year} {1970})}\BibitemShut {NoStop}%
\bibitem [{\citenamefont {Turrell}\ \emph {et~al.}(2015)\citenamefont
  {Turrell}, \citenamefont {Sherlock},\ and\ \citenamefont
  {Rose}}]{Turrell2015}%
  \BibitemOpen
  \bibfield  {author} {\bibinfo {author} {\bibfnamefont {A.~E.}\ \bibnamefont
  {Turrell}}, \bibinfo {author} {\bibfnamefont {M.}~\bibnamefont {Sherlock}}, \
  and\ \bibinfo {author} {\bibfnamefont {S.~J.}\ \bibnamefont {Rose}},\
  }\href@noop {} {\bibfield  {journal} {\bibinfo  {journal} {Journal of
  Computational Physics}\ }\textbf {\bibinfo {volume} {299}},\ \bibinfo {pages}
  {144} (\bibinfo {year} {2015})}\BibitemShut {NoStop}%
\bibitem [{\citenamefont {Casanova}\ \emph {et~al.}(1991)\citenamefont
  {Casanova}, \citenamefont {Larroche},\ and\ \citenamefont
  {Matte}}]{Casanova1991}%
  \BibitemOpen
  \bibfield  {author} {\bibinfo {author} {\bibfnamefont {M.}~\bibnamefont
  {Casanova}}, \bibinfo {author} {\bibfnamefont {O.}~\bibnamefont {Larroche}},
  \ and\ \bibinfo {author} {\bibfnamefont {J.-P.}\ \bibnamefont {Matte}},\
  }\href@noop {} {\bibfield  {journal} {\bibinfo  {journal} {Physical Review
  Letters}\ }\textbf {\bibinfo {volume} {67}},\ \bibinfo {pages} {2143}
  (\bibinfo {year} {1991})}\BibitemShut {NoStop}%
\bibitem [{\citenamefont {Dervieux}\ \emph {et~al.}(2015)\citenamefont
  {Dervieux}, \citenamefont {Loupias}, \citenamefont {Baton}, \citenamefont
  {Lecherbourg}, \citenamefont {Glize}, \citenamefont {Rousseaux},
  \citenamefont {Reverdin}, \citenamefont {Gremillet}, \citenamefont
  {Blancard}, \citenamefont {Silvert}, \citenamefont {Pain}, \citenamefont
  {Brown}, \citenamefont {Allan}, \citenamefont {Hill}, \citenamefont
  {Hoarty},\ and\ \citenamefont {Renaudin}}]{Dervieux2015}%
  \BibitemOpen
  \bibfield  {author} {\bibinfo {author} {\bibfnamefont {V.}~\bibnamefont
  {Dervieux}}, \bibinfo {author} {\bibfnamefont {B.}~\bibnamefont {Loupias}},
  \bibinfo {author} {\bibfnamefont {S.}~\bibnamefont {Baton}}, \bibinfo
  {author} {\bibfnamefont {L.}~\bibnamefont {Lecherbourg}}, \bibinfo {author}
  {\bibfnamefont {K.}~\bibnamefont {Glize}}, \bibinfo {author} {\bibfnamefont
  {C.}~\bibnamefont {Rousseaux}}, \bibinfo {author} {\bibfnamefont
  {C.}~\bibnamefont {Reverdin}}, \bibinfo {author} {\bibfnamefont
  {L.}~\bibnamefont {Gremillet}}, \bibinfo {author} {\bibfnamefont
  {C.}~\bibnamefont {Blancard}}, \bibinfo {author} {\bibfnamefont
  {V.}~\bibnamefont {Silvert}}, \bibinfo {author} {\bibfnamefont {J.~C.}\
  \bibnamefont {Pain}}, \bibinfo {author} {\bibfnamefont {C.~R.~D.}\
  \bibnamefont {Brown}}, \bibinfo {author} {\bibfnamefont {P.}~\bibnamefont
  {Allan}}, \bibinfo {author} {\bibfnamefont {M.~P.}\ \bibnamefont {Hill}},
  \bibinfo {author} {\bibfnamefont {D.~J.}\ \bibnamefont {Hoarty}}, \ and\
  \bibinfo {author} {\bibfnamefont {P.}~\bibnamefont {Renaudin}},\ }\href@noop
  {} {\bibfield  {journal} {\bibinfo  {journal} {High Energy Density Physics}\
  }\textbf {\bibinfo {volume} {16}},\ \bibinfo {pages} {12} (\bibinfo {year}
  {2015})}\BibitemShut {NoStop}%
\bibitem [{\citenamefont {Decyk}(1987)}]{Decyk:1987vt}%
  \BibitemOpen
  \bibfield  {author} {\bibinfo {author} {\bibfnamefont {V.~K.}\ \bibnamefont
  {Decyk}},\ }\href@noop {} {\bibfield  {journal} {\bibinfo  {journal} {Report
  PPG-1057, Invited paper presented at the International Conference on Plasma
  Physics, Kiev, USSR}\ } (\bibinfo {year} {1987})}\BibitemShut {NoStop}%
\end{thebibliography}%
\bibliographystyle{apsrev4-1.bst}

\end{document}